\documentclass[superscriptaddress,preprintnumbers,amssymb,twocolumn,nofootinbib]{revtex4}
\pdfoutput=1
\usepackage{graphicx}
\usepackage{epsfig}
\usepackage{amsmath}
\usepackage{amsfonts}
\usepackage{amssymb}

\usepackage{braket}
\usepackage{cancel}
\usepackage{pstricks}
\usepackage{color}
\usepackage{feynmf}

\usepackage[normalem]{ulem}

\def\lsim{\:\raisebox{-0.5ex}{$\stackrel{\textstyle<}{\sim}$}\:}
\def\gsim{\:\raisebox{-0.5ex}{$\stackrel{\textstyle>}{\sim}$}\:}

\def\missET {{\not\!\! E_T}}

\usepackage{epstopdf}

\begin{document}

\title {Phenomenology of TeV-scale scalar Leptoquarks
in the EFT}
%
\author{Shaouly Bar-Shalom}
\email{shaouly@physics.technion.ac.il}
\author{Jonathan Cohen}
\email{jcohen@tx.technion.ac.il}
\affiliation{Physics Department, Technion-Institute of Technology, Haifa 32000, Israel}
\author{Amarjit Soni}
\email{adlersoni@gmail.com}
\affiliation{Physics Department, Brookhaven National Laboratory, Upton, NY 11973, USA}
\author{Jose Wudka}
\email{jose.wudka@ucr.edu}
\affiliation{Physics Department, University of California, Riverside, CA 92521, USA}

\date{\today}

\begin{abstract}
We examine new aspects of leptoquark (LQ) phenomenology using effective field theory (EFT).
We construct a complete set of leading effective operators involving
SU(2) singlets scalar LQ and the SM fields up to dimension six.
We show that,
while the renormalizable LQ-lepton-quark
interaction Lagrangian can address the persistent hints for physics beyond
the Standard Model in the B-decays $\bar B \to D^{(*)} \tau \bar\nu$,
$\bar B \to \bar K \ell^+ \ell^-$ and
in the measured anomalous magnetic moment of the muon,
the LQ higher dimensional effective operators
may lead to new interesting effects associated with
lepton number violation.
These include the generation of one-loop and two-loops sub-eV Majorana
neutrino masses, mediation of neutrinoless double-$\beta$ decay and novel
LQ collider signals. For the latter, we focus
on 3rd generation LQ ($\phi_3$) in a framework with an approximate
$Z_3$ generation symmetry, and show that one class of the
dimension five LQ operators
may give rise to a striking asymmetric
same-charge $\phi_3 \phi_3$ pair-production signal, which
leads to low background same-sign leptons signals at the LHC.
For example, with $M_{\phi_3} \sim 1$ TeV and a new physics scale
of $\Lambda \sim 5$ TeV, we expect at the 13 TeV LHC with
an integrated luminosity of 300 fb$^{-1}$, about
$5000$ positively charged $\tau^+ \tau^+$
events via $\boldsymbol{pp \to \phi_3 \phi_3  \to \tau^+ \tau^+  + 2 \cdot j_b}$ ($j_b$=b-jet), about $500$ negatively charged $\tau^- \tau^-$ events with a signature $\boldsymbol{pp \to \phi_3 \phi_3 \to \tau^- \tau^- + 4 \cdot j + 2 \cdot j_b}$ ($j$=light jet) and about $50$ positively charged $\ell^+ \ell^+$ events
via $\boldsymbol{pp \to \ell^+ \ell^+  + 2 \cdot j_b + \missET}$
for any of the three charged leptons,
$\ell^+ \ell^+ = e^+ e^+,\mu^+ \mu^+, \tau^+ \tau^+$.
It is interesting to note that, in
the LQ EFT framework, the expected
same-sign lepton signals have a rate
which is several times larger than the
QCD LQ-mediated
opposite-sign leptons signals,
$gg,q \bar q \to \phi_3 \phi_3^* \to \ell^+ \ell^- +X$.
We also consider the same-sign charged lepton signals in the LQ EFT framework at higher energy hadron colliders such as a 27 TeV HE-LHC and a 100 TeV FCC-hh.
\end{abstract}


\maketitle

\section{Introduction \label{sec1}}

The electroweak (EW) and strong interactions of the SM have been very
successfully tested at the low-energy (GeV-scale) and high-energy (EW-scale) frontiers as well as
in precision measurements \cite{PDG}. However, despite the impressive success of the SM
at sub-TeV energies, it is widely believed that it is an effective low-energy
framework of a more complete UV theory that should address the
experimental and theoretical indications for new physics
beyond the SM (BSM),
such as the indirect detection of dark matter and dark energy, the measurements of neutrino masses,
the flavor and hierarchy problems residing in the SM's scalar sector and the long
sought higher symmetry which unifies the fundamental forces.

The scale of the new physics (NP) that may shed light
on these fundamental questions in particle
physics and address the deficiencies of the SM
might be beyond the reach of present and future high-energy colliders.
Nonetheless, the underlying UV theory may
contain new particles with masses spanning over
many orders of magnitudes, similar to the hierarchical mass pattern
observed in nature and embedded in the SM.
Indeed, although direct searches at high-energy colliders have not yet led to a discovery
of new heavy particles, there have been intriguing and persistent hints
in the past several years in favor of new TeV-scale degrees of freedom from
measured anomalies associated with possible violations of lepton universality in B-decays:
$\bar B \to D^{(*)} \tau \bar\nu$ \cite{BtoD1,BtoD2,BtoD3} and
$\bar B \to \bar K \ell^+ \ell^-$ \cite{BtoK}, as well
as in the anomalous magnetic moment of the muon \cite{muong}.

Out of these three anomalies, the most striking and least expected  is the anomalous
enhanced  $\bar B \to D^{(*)} \tau \bar\nu$ rate measured by BaBar \cite{BtoD1},
Belle \cite{BtoD2} and LHCb \cite{BtoD3} (a $\sim 4 \sigma$ effect). In the SM this decay occurs at tree-level and is mediated by the $Wcb$ charged current coupling,
so that the measured deviation requires a relatively
large tree-level NP contribution near the TeV scale to compete with the ``classic" SM tree-level
diagram. Promising candidates that address this large
effect in $\bar B \to D^{(*)} \tau \bar\nu$ are TeV-scale leptoquarks (LQ's);
in addition to this phenomenological role, these particles also appear naturally in theories
that address some of the most fundamental questions in particle physics (see
\cite{1603.04993} and references therein) such as grand unification \cite{LQ1} and compositeness
\cite{LQ3}, where they can also arise as pseudo-Nambu Goldstone bosons
\cite{LQ31} and lead to interesting collider signals
\cite{1803.05962,1808.02063}.
They are also involved
in models for neutrino masses \cite{LQ4}.
In some cases, the effects of scalar LQ are similar to
that of the scalar partners of the quarks in R-parity violating
supersymmetry models \cite{Neubert,soniRPV}, which can
have similar couplings to quark-lepton pairs.

Given their theoretical appeal, and their potential role in addressing the $B$ anomalies,
it is of interest to study LQ phenomenology within the context of BSM physics.
That is, allowing for the presence of excitations heavier than the LQs.
This we shall do using an effective field theory, which will include the LQs as
(relatively) low-energy excitations, and the effective interactions generated
by heavier physics of scale $ \Lambda $. Indeed, the mere presence of the
TeV-scale renormalizable LQ framework (e.g., its Yukawa-like couplings
to a quark-lepton pair which is being used in order to address the B-anomalies)
suggests that this EFT higher-dimensional expansion is well defined and
can be constructed in principle to any order (see section \ref{sec3}).
If the NP scale $ \Lambda $ is much higher than the multi TeV-scale,
i.e., $\Lambda \gg 10$ TeV, then the effects of these higher-dimensional
LQ effective interactions will be negligible.
On the other hand, the purpose of this paper is to
investigate the extent to which the LHC can probe
physics beyond the LQ mass;
we will see that, if $\Lambda \sim 5 -15$ TeV, then the
higher dimensional LQ effective interactions can produce unique
collider signatures
that may be observable at the LHC, and, in some cases, at rates that
are {\em higher} than for the usual channels. We will also see that
the physics at scale $ \Lambda$, responsible for the effective LQ
interactions, is also intimately connected with various possible mechanism
of neutrino mass generation, so that a study of LQ phenomenology at the LHC
can provide also information about the neutrino sector.

In this work we will concentrate on the
study of the interactions and phenomenology of
 TeV-scale scalar LQs, which are SU(2) singlets and transform
either as a right-handed down-type quark,$^{1}$
$\phi(3,1,-\frac{1}{3})$,
or as a right-handed up-type quark, $\phi(3,1,\frac{2}{3})$,
under the SM gauge group; since the BSM effects of both types of LQ have similar characteristics, in the bulk
of the paper we will explore the effects and underlying
physics of the down-type LQ, and towards the end of the paper we will shortly address the underlying physics and effects that are expected for an up-type LQ.

We  construct the complete set of  effective operators up to dimension six that involve the LQs
and SM fields, and use this LQ EFT framework to demonstrate the impact of heavy physics
on $\phi$ collider phenomenology, and
on low-energy lepton number violating (LNV) phenomena such
as Majorana neutrino masses and neutrino-less double beta
decay. This model-independent formalism provides a broader and a more
reliable view of the expected physics associated with TeV-scale LQs,
and  lays the ground
for further investigations of $\phi$-related phenomenology at high-energy colliders.
For example, we find that the higher dimensional LQ interactions in the EFT framework may lead to very interesting, essentially background free, same-sign lepton signals at the LHC and/or at future colliders.

Lastly, we want to stress that while our starting motivation for this
work was the B-anomalies, the confirmation of the anomalies is
not needed for our work to have merit.
Indeed, as was mentioned above, leptoquarks dynamics may be linked to
well motivated extensions of the SM, such as composite theories
and R-parity violating supersymmetry and, in particular, they play
an important role in Grand Unified theories.

The paper is organized as follows: in the following section we summarize the
effects of the renormalizable LQ interaction Lagrangian
${\cal L}_{\phi SM}$;
in section \ref{sec2} we review
the LHC phenomenology of the scalar LQ in the $\phi SM$ framework and
in section \ref{sec3} we construct the effective theory beyond
${\cal L}_{\phi SM}$, listing all the higher-dimensional effective
operators involving the down-type LQ $\phi(3,1,-\frac{1}{3})$
up to dimension six. In section \ref{sec4} we study the
$\Delta L =2$ low-energy effects associated with the dimension
five operators and in section \ref{sec5} we explore the leading
signals of the down-type and up-type
LQ, $\phi(3,1,-\frac{1}{3})$ and $\phi(3,1,\frac{2}{3})$,
in the EFT framework at the 13 TeV LHC as well as
at higher energy (27 and 100 TeV) hadron colliders.
In Section \ref{sec6} we summarize and in the appendix we list all
dimension six
operators for the down-type LQ.

\medskip

\section{Renormalizable LQ interactions}

We define the renormalizable extension of the SM
which contains the LQ as:
\begin{eqnarray}
{\cal L}_{\phi SM} = {\cal L}_{SM} + {\cal L}_{Y,\phi} + {\cal L}_{H,\phi} ~, \label{phiSM}
\end{eqnarray}
where, for the down-type LQ $\phi(3,1,-\frac{1}{3})$, the Yukawa-like and scalar interaction pieces are:
\begin{widetext}
\begin{eqnarray}
{\cal L}_{Y,\phi} &=& y_{q \ell}^L \bar q^c i \tau_2 \ell \phi^* +
y_{u e}^R \bar u^c e \phi^* + y_{qq}^L \bar q^c i \tau_2 q \phi
+ y_{u d}^R \bar u^c d \phi + \mbox{H.c.} ~, \label{phiSMY} \\
{\cal L}_{H,\phi} &=& |D_{\mu} \phi|^2 - M_\phi^2 |\phi|^2 + \lambda_\phi |\phi|^4 +
\lambda_{\phi H} |\phi|^2 |H|^2 ~, \label{phiSMH}
\end{eqnarray}
\end{widetext}
with $q$ and $\ell$ the SU(2) left-handed quark and lepton doublets, respectively, while
$u,d,e$ are the right-handed SU(2) singlets; also, $\psi^c = C \bar\psi^T$.

\footnotetext[1]{In our notation $X(c,w,y)$, indicates that particle $X$ transforms under SU(3) representation $c$, SU(2) dimension $w$ and carries hypercharge $y$.}

A few comments are in order regarding the
$\phi SM$ Lagrangian defined in Eqs.~\ref{phiSM}-\ref{phiSMH}:
\begin{itemize}
\item The last two Yukawa-like $\phi$-quark-quark terms of
${\cal L}_{Y,\phi}$ in Eq.~\ref{phiSMY} violate
Baryon number and can potentially mediate proton decay (see e.g., \cite{pdecay}).
The Yukawa-like LQ couplings involving the 1st and 2nd generations are then either vanishingly small (i.e., $(y_{qq}^L)_{ij},(y_{ud}^R)_{ij} \to 0$ for $i,j \neq 3$) or are
forbidden, e.g., by means of a symmetry.
\item The first two Yukawa-like $\phi$-quark-lepton terms
of ${\cal L}_{Y,\phi}$ in Eq.~\ref{phiSMY} (i.e.,
$ \propto y_{q \ell}^L,y_{ue}^R$) can address
the enhanced rate measured
in the tree-level $\bar B \to D^{(*)} \tau \bar\nu$ decay as well as
the 1-loop anomalies observed in
$\bar B \to \bar K \ell^+ \ell^-$ and the muon magnetic moment
\cite{Neubert,ref13,1408.1627,1505.05164,ref17,1612.07757}, when
$M_\phi \sim {\cal O}(1)$ TeV and couplings
$y_{q \ell}^L,y_{ue}^R \sim {\cal O}(0.1 -1)$.
It should be noted, though, that
these down-type LQ $\phi$-quark-lepton
interactions are not sufficient for a simultaneous explanation of
all these anomalies \cite{1608.07583,1703.09226,1806.05689,1806.07403,1807.02068,1808.08179,1810.11588}.

\item The LQ - Higgs interaction term $\propto \lambda_{\phi H}$ in Eq.~\ref{phiSMH}
may play an important role in stabilizing the EW vacuum \cite{vacuumS}.
\item As will be discussed below, within the renormalizable
$\phi SM$ framework, ${\cal L}_{\phi SM}$,
LQ phenomenology and leading signals at the LHC are completely determined
by the two Yukawa-like parameters $y_{q \ell}^L,y_{ue}^R$ and the LQ mass
$M_{\phi}$ (ignoring the baryon number violating couplings).
\end{itemize}

\section{Phenomenology of scalar leptoquarks in the $\phi SM$ framework \label{sec2}}

In the limit $y_{q \ell}^L,y_{ue}^R \to 0$ the only production channels of a scalar LQ at the LHC
 are the tree-level QCD $\phi \phi^*$ pair-production via
$gg \to \phi \phi^*$ and the s-channel gluon exchange in
$q \bar q$-fusion $q \bar q \to \phi \phi^*$, see e.g.,
\cite{9610408,9704322,0411038,1406.4831,1506.07369,1801.04253,1801.07641,1810.10017,1811.03561}.
The corresponding typical $\phi \phi^*$ pair-production
cross-section at the 13 TeV LHC is
$\sigma_{\phi \phi^*} \sim 5(0.01)$ fb for
$M_\phi \sim 1(2)$ TeV \cite{1801.07641}.
Turning on the Yukawa-like $\phi$-quark-lepton interactions in Eq.~\ref{phiSMY} adds
another tree-level t-channel lepton exchange diagram to $q \bar q \to \phi \phi^*$,
which, however, is subdominant. Thus, LQ pair-production at the LHC
is essentially independent of its Yukawa-like couplings to a quark-lepton pair.

On the other hand, with sizable $y_{q \ell}^L,y_{ue}^R$ Yukawa terms,
the LQ $\phi$ can also be singly produced
at tree-level by the quark-gluon fusion processes $q g \to \phi \ell$;
for $\phi = (3,1,-\frac{1}{3})$ there are two
production channels $u g \to \phi \ell_i$ and $d g \to \phi \nu_i$,
where $i=1,2,3$ is a generation index and both channels
include two diagrams: an s-channel $q$-exchange and t-channel $\phi$-exchange.
The single LQ production channel
is in fact dominant if $\phi$ has ${\cal O}(1)$ Yukawa-like couplings to the 1st
generation quarks: $\sigma_\phi^{\rm single} = \sigma(q g \to \phi \ell) \propto y_{q \ell}^2$ (here $q=u,d$ and $\ell = e,\nu_e$),
and
with $y_{q \ell} \sim {\cal O}(1)$ one obtains
$\sigma_\phi^{\rm single}(pp_{(u g)} \to \phi e) \sim 100(2)$ fb and
$\sigma_\phi^{\rm single}(pp_{(d g)} \to \phi \nu_e) \sim 50(0.5)$ fb
for $M_\phi = 1(2)$ TeV, see e.g., \cite{1801.07641}.

The search for LQ is then performed assuming two distinct LQ decay channels
that correspond to its two Yukawa-like interactions in the $\phi SM$: $\phi \to e_i j$
and $\phi \to \nu j$, with
$\Gamma(\phi \to e_i j/\nu j) \sim |y|^2 m_{\phi}/16 \pi$, where
$y$ is the corresponding $\phi$-lepton-quark coupling, and
the quark and lepton masses are neglected.
Thus, the overall LQ signatures at the LHC
contain either two leptons and two jets with large
transverse momentum,
$e_i^+ e_j^- j j$ and/or $e_i jj + {\rm missing}~ E_T$, when the LQ are pair-produced
\cite{1612.01190,1703.03995,CMSEXO017009,CMSEXO17016,1809.05558,1803.02864,1808.05082},
or two leptons and a jet with large
transverse momentum,
$e_i^+ e_j^- j$ and $e_i j + {\rm missing}~ E_T$, when the LQ is singly produced.

Indeed, searches for 1st and 2nd generations LQ
pair-production (i.e. for LQ with
couplings only to quark-lepton pairs of the 1st and 2nd generations)
yield stronger bounds than the ones for 3rd generation LQ,
since the detector sensitivity to the different flavors of high-$p_T$
leptons and quarks varies. In addition,
these bounds strongly depend on the LQ decay pattern, i.e., branching ratios to the
different quark-lepton pairs. For example, the current bounds on the mass of a
1st(2nd) generation LQ assuming $pp \to \phi \phi^* \to e^+ e^-/\mu^+ \mu^- + jj$ and
$BR(\phi \to e/\mu + j) \sim 1$
is $M_\phi \gsim 1.5$ TeV \cite{CMSEXO017009,1808.05082}.

Third generation LQ are particularly motivated, due to their
potential role in explaining the observed anomalies in B-physics discussed above,
but also on more general aspects concerning the underlying
UV physics, e.g., the dynamical generation of fermion masses in composite
scenarios \cite{3rdLQ}.
Recent searches for a pair-produced 3rd generation scalar LQ, decaying via $\phi \to t \tau, b \nu_\tau$ and/or $\phi \to b \tau$,
have yielded weaker bounds: $M_\phi \gsim 1$ TeV \cite{1612.01190,1703.03995,CMSEXO17016,1809.05558,1803.02864,1010.3962}.
On the other hand, the bound on the mass of a $\phi(3,1,-\frac{1}{3})$ that couples exclusively to a top-muon pair (and can, therefore, address the
anomalous muon magnetic moment and the anomaly
measured in $\bar B \to \bar K \ell^+ \ell^-$), obtained in the search for
$pp \to \phi \phi^* \to t \bar t \mu^+ \mu^-$, is
$M_\phi \gsim 1.4$ TeV \cite{1809.05558}, i.e., comparable
to the lower limit on the mass of a 1st and 2nd generation LQ.
Furthermore, a search for a singly produced 3rd generation scalar LQ
which decays exclusively via $\phi \to b \tau$
has also been performed recently by CMS; they exclude such a LQ
up to a mass of 740 GeV \cite{1806.03472}.

Finally, another important LQ-mediated signal
is the t-channel LQ exchange in the Drell-Yan lepton pair-production process $q \bar q \to \ell^+ \ell^-$.
In particular, this channel becomes important in the large LQ-lepton-quark coupling regime, since
the corresponding cross-section scales as
$\sigma(q \bar q \to  \ell^+ \ell^-) \propto y_{q \ell}^4$,
thus providing a complimentary sensitivity to the LQ dynamics as the LHC
\cite{1603.04993,1810.10017,1811.03561,1409.2372,1609.07138,1704.09015,1901.10480}; in particular
yielding better access to larger LQ masses where the QCD on-shell LQ pair production channel
is suppressed.

\section{EFT beyond the $\phi SM$ framework \label{sec3}}

In this section we focus on the EFT extension
of the renormalizable Lagrangian in Eqs.~\ref{phiSM}-\ref{phiSMH},
for the down-type LQ $\phi(3,1,-\frac{1}{3})$.
The effects of the NP which
underlies the $\phi$SM framework in Eqs.~\ref{phiSM}-\ref{phiSMH}
can be parameterized by a series of effective operators
$O_i$, which are constructed using the $\phi$SM fields and whose coefficients are
suppressed by inverse powers of the NP scale $\Lambda$,
\begin{eqnarray}
{\cal L} = {\cal L}_{\phi SM} + \sum_{n=5}^\infty
\frac{1}{\Lambda^{n-4}} \sum_i f_i O_i^{(n)} \label{EFT1}~,
\end{eqnarray}
where $n$ is the mass dimension of $O_i^{(n)}$
and we assume decoupling and weakly-coupled heavy NP, so that
$n$ equals the canonical dimension. The dominating NP effects
are then expected to be generated by
contributing operators
with the lowest dimension ($n$ value)
that can be generated at tree-level in the underlying theory.
\begin{figure}[htb]
\begin{center}
\includegraphics[scale=0.5]{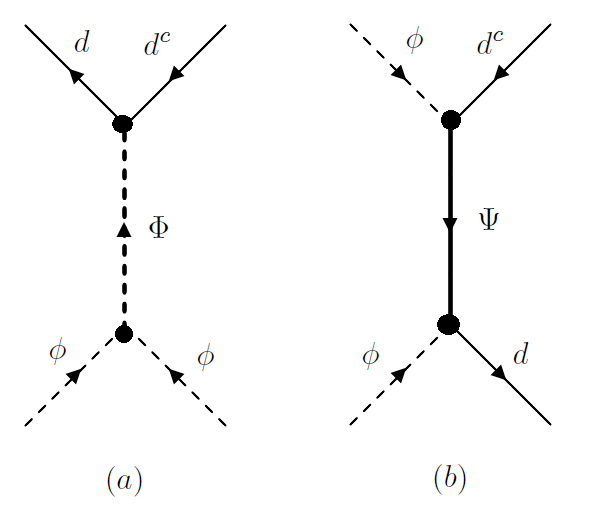}
\end{center}
\vspace{-0.8cm}
\caption{Tree-level graphs in the underlying heavy theory that generate the
dimension five effective operator $\bar d d^c \phi^2$. $\Phi$ and $\Psi$
stand for a heavy scalar and heavy fermion, respectively, with quantum numbers
$\Phi(6,1,-\frac{2}{3})$ and $\Psi(1,1,0)$ or $\Psi(8,1,0)$ (see text).}
\label{figd2p2}
\end{figure}
\begin{figure}[htb]
\begin{center}
\includegraphics[scale=0.5]{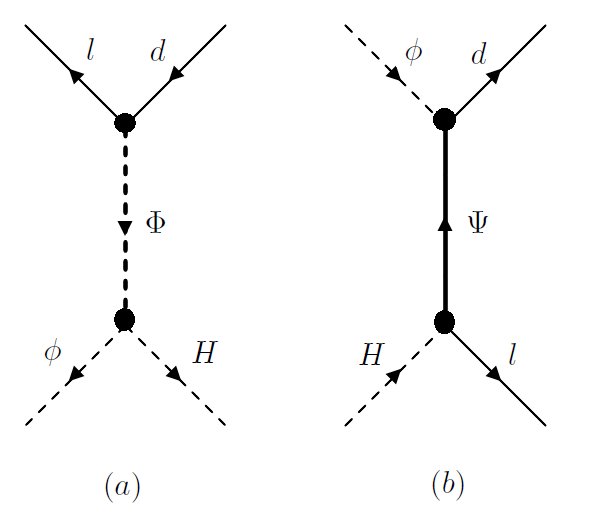}
\end{center}
\vspace{-0.8cm}
\caption{Tree-level graphs in the underlying heavy theory that generate the
dimension five effective operator $\bar\ell d \tilde H \phi^*$. $\Phi$ and $\Psi$
stand for a heavy scalar and heavy fermion, respectively, with quantum numbers
$\Phi(3,2,\frac{1}{6})$ and $\Psi(1,1,0)$, $\Psi(1,3,0)$ or
$\Psi(3,2,-\frac{5}{6})$ (see text).}
\label{figldhp}
\end{figure}

Before listing the specific form of the higher dimension operators, $O_i^{(n)}$,
it is useful to denote their generic structure in the form
\begin{eqnarray}
O_i^{(n)} \in \phi^a H^b \psi^c D^d ~, \label{gen1}
\end{eqnarray}
where $a,b,c,d$ are integers representing the multiplicity
of the corresponding factors: $O_i^{(n)}$ contains $a$ LQ fields $\phi$ or $\phi^*$, $b$ Higgs fields $H$ or $\tilde H$, $c$ fermionic fields $\psi$ and $d$ covariant derivatives $D$. Group contractions and which fields are acted on by the derivatives are not specified.

We find that there are only two possible
dimension-five operators involving the LQ $\Phi(3,1,-\frac{1}{3})$
and the SM fields -- both violating lepton number
by two units.
To see that, note that the dimension-five
operators with $c=0$ in Eq. \ref{gen1}
are all absent because of gauge invariance. Furthermore,
operators of the form $\phi^2 \psi^2$ must contain the fermion bilinear $\bar\psi_L \psi_R$, so that only a single gauge invariant dimension five operator of this form survives
(with two possible SU(3) color contractions which are not specified):
\begin{eqnarray}
O_{d^2 \phi^2}^{(5)} = \bar d d^c \phi^2 \label{d2p2dim5}~,
\end{eqnarray}
which violates lepton number by two units.

The diagrams that can generate the dimension five operator $\bar d d^c \phi^2$
at tree-level in the underlying heavy theory are depicted
in Fig.~\ref{figd2p2}; the corresponding heavy NP
must contain a heavy scalar
$\Phi(6,1,-\frac{2}{3})$ and/or the heavy fermions
$\Psi(1,1,0)$, $\Psi(8,1,0)$.

Dimension five operators of the class $\phi \psi^2 D$ can be shown to be equivalent to operators
without a derivative using integration by parts and, therefore, can be ignored. Thus, the remaining class of
dimension five operators is of the form
$\phi \psi^2 H$ and, therefore, must also contain the fermion bilinear $\bar\psi_L \psi_R$. The only gauge invariant  operator of this form, which also violates lepton number by two units is:
\begin{eqnarray}
O_{\ell d \phi H}^{(5)} = \bar\ell d \tilde H \phi^* \label{ldhpdim5}~.
\end{eqnarray}

The heavy physics generating
this operator at tree-level must contain a heavy scalar
$\Phi(3,2,\frac{1}{6})$ and/or the heavy fermions
$\Psi(1,1,0)$, $\psi(1,3,0)$ or
$\Psi(3,2,-\frac{5}{6})$, see Fig.~\ref{figldhp}.

We recall that there is also a unique
dimension five operator that can be constructed using
the SM fields only; the so called Weinberg operator \cite{Weinberg}:
\begin{eqnarray}
O_{W}^{(5)} = \bar\ell^c \tilde H^\star
\tilde H^\dagger \ell  \label{h2l2dim5}~,
\end{eqnarray}
that can be generated in the underlying theory
at tree-level by an exchange of a heavy scalar
$\Phi(1,3,0)$ and/or the heavy fermions
$\Psi(1,1,0)$, $\Psi(1,3,0)$.

Therefore, the overall dimension five effective
operator extension of ${\cal L}_{\phi SM}$ is:
\begin{widetext}
\begin{eqnarray}
\Delta {\cal L}_{\phi SM}^{(5)} =
\frac{f_{W}}{\Lambda_{W}} \bar\ell^c \tilde H^\star  \tilde H^\dagger \ell  +
\frac{f_{\ell d \phi H}}{\Lambda_{\ell d \phi H}} \bar\ell d \tilde H \phi^* +
\frac{f_{d^2 \phi^2}}{\Lambda_{d^2 \phi^2}} \bar d d^c \phi^2 + \mbox{H.c.} \label{dim5all}~,
\end{eqnarray}
\end{widetext}
where we have kept a general notation assigning each of these operators their own
effective scale. Note, for example, that the heavy fermionic state $\Psi(1,1,0)$ can generate all
three dimension five operators in Eq.~\ref{dim5all},
in which case they will have a common
scale. On the other hand, as we will see below,
the effective scale, $\sim f/\Lambda$,
of the
Weinberg operator $\bar\ell^c \tilde H^\star  \tilde H^\dagger \ell$ and
the operator $\bar\ell d \tilde H \phi^*$ must
be considerably suppressed in order to obtain sub-eV
Majorana neutrino masses.
This leaves us with a single viable dimension five operator,
$\bar d d^c \phi^2$, which can generate a sub-eV neutrino mass at two-loops (see next section) with a
scale low enough for it to be relevant
for collider LQ phenomenology.

In the appendix we construct the complete set of
the dimension six operators involving the down-type
scalar LQ $\phi(3,1,-\frac{1}{3})$ and the
SM fields.$^{2}$

\footnotetext[2]{We have used the Mathematica notebook of \cite{1512.03433} to validate the EFT extension
of ${\cal L}_{\phi SM}$ which is presented in this work.}

\section{The dimension five operators and low energy $\Delta L =2$ effects \label{sec4}}

As mentioned earlier, while the
$\phi$SM renormalizable interaction Lagrangian,
${\cal L}_{\phi SM}$, can address
the BSM effects associated with
the current B-physics anomalies, other aspects of NP associated with
LNV require new
higher-dimensional effective interactions of the LQ with the SM fields.
In particular, the dimension five operators in Eq.~\ref{dim5all}
violate lepton number by two units and can, therefore,  generate Majorana neutrino masses, mediate
neutrinoless double beta decay and also give rise to interesting same-sign lepton signals at the LHC.

In this section we investigate in more detail the low energy $\Delta L =2$ effects associated
with these operators, while in the next section we discuss
the potential $\Delta L =2$ collider signals.

\subsection{Majorana Neutrino masses \label{sec42}}

As is well known, the dimension five Weinberg operator
$\bar\ell^c \tilde H^\star  \tilde H^\dagger \ell$ can generate
a tree-level Majorana neutrino mass through the type I
(if it is generated by the exchange of the heavy fermion $\Psi(1,1,0)$)
and/or type III (if it is generated by $\Psi(1,3,0)$) seesaw mechanisms.
In either case, the resulting Majorana neutrino mass is:
\begin{eqnarray}
m_\nu (\Lambda) \sim f_W \cdot \frac{v^2}{\Lambda_W} ~,
\end{eqnarray}
where $v$ is the Higgs Vacuum Expectation Value (VEV) and $f_{W}$ and $\Lambda_W$ are the Wilson coefficient and NP scale of the Weinberg operator (see Eq.~\ref{dim5all}).

Therefore, there are two extreme cases for generating
$m_\nu \lsim 1$ eV from $O_W^{(5)}$:
either $\Lambda_{W} \sim {\cal O}(10^{14})$ GeV and
$f_{W} \sim {\cal O}(1)$ or, if the NP scale
is at the TeV range, i.e., $\Lambda_{W} \sim {\cal O}(1)$ TeV, then
$f_{W} \sim {\cal O}(10^{-11})$.
In both cases the effect of the Weinberg operator at TeV-scale
energies is negligible.
\begin{figure}[htb]
\begin{center}
\includegraphics[scale=0.5]{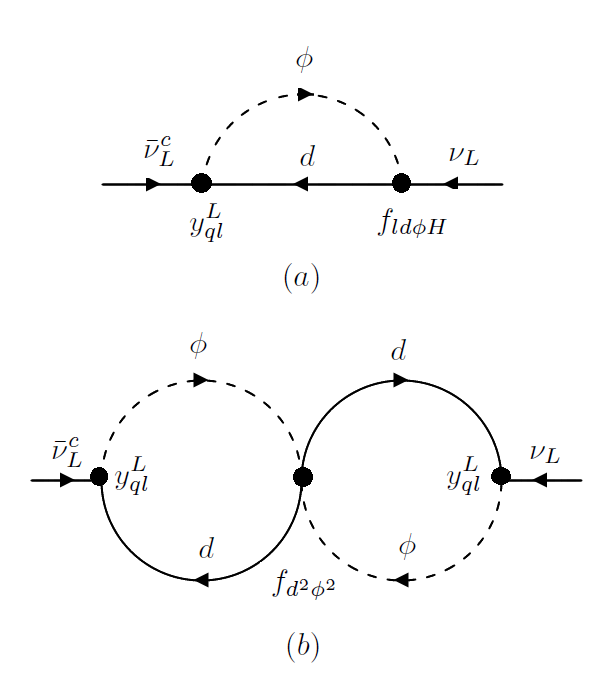}
\end{center}
\vspace{-0.8cm}
\caption{The one-loop and two-loops
diagrams (a) and (b) which generates a Majorana mass term with
the Yukawa-like LQ-quark-lepton interaction
($ \propto y_{q\ell}^L$) and the dimension five operators
$\bar\ell d \tilde H \phi^*$ and $\bar d d^c  \phi^2$
(with the coupling strength $f_{\ell d \phi H}$ and $f_{d^2 \phi^2}$, respectively).
See also text.}
\label{Mneu}
\end{figure}

The operators $\bar\ell d \tilde H \phi^*$ and $\bar d d^c  \phi^2$
can also generate a Majorana neutrino mass term
at 1-loop and 2-loops order, respectively, via the diagrams
depicted in Fig.~\ref{Mneu}. In particular, this
involves insertions of the dimension five coupling strengths
$f_{\ell d \phi H}$ and $f_{d^2 \phi^2}$
as well as the Yukawa-like LQ-quark-lepton renormalizable interaction
$\propto y_{q \ell}^L$ of the $\phi$SM Lagrangian in Eq.~\ref{phiSMY}.
For the $\bar\ell d \tilde H \phi^*$ case, the resulting 1-loop
Majorana mass is:$^{3}$\footnotetext[3]{See also Eq.26 in \cite{soniRPV} for an analogous
down-quark - down-squark 1-loop Majorana mass term in R-parity violating
Supersymmetry.}
\begin{eqnarray}
m_\nu (\Lambda) \sim \frac{3 m_d}{16 \pi^2} \frac{f \cdot y_{q \ell}^L }{\sqrt{2}}
\frac{v}{\Lambda} \ln\left(\frac{\Lambda^2}{M_{\phi}^2}\right) ~,
\end{eqnarray}
where $\Lambda = \Lambda_{\ell d \phi H}$ and
$f=f_{\ell d \phi H}$ are the NP scale and Wilson coefficient
of the dimension five operator $\bar\ell d \tilde H \phi^*$;
$m_d$ is the mass of the down-quark
in the loop and $M_\phi$ is
the leptoquark mass. Thus,
setting e.g., $\Lambda= 5$ TeV and $M_{\phi}=1$ TeV, we obtain:
\begin{eqnarray}
\frac{m_\nu (\Lambda = 5~{\rm TeV})}{f \cdot y_{q \ell}^L}
\sim 10^{-3} \cdot m_d ~,
\end{eqnarray}
so that, for $f \cdot y_{q \ell}^L \sim {\cal O}(1)$,
the resulting Majorana mass is
$m_\nu \sim {\cal O}({\rm KeV})$ for
$m_d \sim {\cal O}({\rm MeV})$ (i.e., the d-quark) and
$m_\nu \sim {\cal O}({\rm MeV})$ for
$m_d \sim {\cal O}({\rm GeV})$
(i.e., the b-quark).
Thus, in order to obtain sub-eV Majorana neutrino masses
when $ \Lambda = {\cal O}(\mbox{TeV})$ we should have
$f \cdot y_{q \ell}^L \lsim {\cal O}(10^{-3})$
for the d-quark loop and $f \cdot y_{q \ell}^L \lsim {\cal O}(10^{-6})$ for
the b-quark loop. In particular, if $\phi$ is a 3rd generation LQ (i.e., having ${\cal O}(1)$ couplings
only to the 3rd generation SM fermions, see next section), then $y_{b \nu}^L \sim {\cal O}(1)$ and,
therefore, the corresponding dimension five coupling strength should be suppressed
to the level $f_{\ell d \phi H} \lsim {\cal O}(10^{-6})$
if $\Lambda_{\ell d \phi H} \sim 5$ TeV,
in order to obtain e.g.,
$m_{\nu_\tau} \lsim 1$ eV
(ignoring off-diagonal generation couplings).
We note that other interesting mechanisms for generating
light Majorana neutrino masses from 1-loop LQ exchanges that are
intimately related to the down-quark mass matrix have been discussed in
\cite{9905340,9909518,0710.5699,1502.05188,1510.08757,1610.02322,1701.08322}.
These studies, however, were based on renormalizable LQ extensions of the SM.

The 2-loop Majorana mass generated by the $\bar d d^c \phi^2$
class of dimension 5 operators  is (see Fig.~\ref{Mneu}):
\begin{eqnarray}
m_\nu (\Lambda) \sim \frac{f \cdot \left( y_{q \ell}^L \right)^2 }{(16 \pi^2)^2}
\frac{3 m_d^2}{\Lambda} \cdot \ln^2\left(\frac{\Lambda^2}{M_{\phi}^2}\right) ~,
\end{eqnarray}
where here $\Lambda = \Lambda_{d^2 \phi^2}$ and
$f=f_{d^2 \phi^2}$ are the NP scale and Wilson coefficient
of the dimension five operator $\bar d d^c \phi^2$.
Thus,
setting again $\Lambda= 5$ TeV and $M_{\phi}=1$ TeV,
we obtain in the 2-loop case:
\begin{eqnarray}
\frac{m_\nu (\Lambda = 5~{\rm TeV})}{f \cdot \left( y_{q \ell}^L \right)^2}
\sim 10^{-4} \cdot \frac{m_d^2}{\rm TeV} ~,
\end{eqnarray}
which, as in the 1-loop case,
depends on the down-quark mass in the loops or, equivalently,
on the LQ generation (defined through its renormalizable couplings
to the quark-lepton pairs, see discussion above).
In particular, here also, it is useful to distinguish between the three cases where
$\phi$ couples to 1st, 2nd or 3rd generation quarks:
\begin{description}
\item{d-quark case ($y_{q \ell}^L = y_{d \nu}^L$ and $f=f_{d^2 \phi^2}$):} \\
In this case the 2-loop neutrino mass is too small,
$m_\nu \sim 10^{-4} ~ {\rm eV}$, when
$f_{d^2 \phi^2} \cdot \left( y_{d \nu}^L \right)^2 \sim {\cal O}(1)$,
so that no useful bound can be set on the scale of the dimension 5 operator
involving the 1st generation down-quarks
$\bar d d^c \phi^2$. Indeed, the collider effects of
this operator, with a scale $\Lambda_{d^2 \phi^2} \sim 5-15$ TeV and
$f_{d^2 \phi^2} \sim {\cal O}(1)$, will be studied in the next sections.
\item{s-quark case ($y_{q \ell}^L = y_{s \nu}^L$ and $f=f_{s^2 \phi^2}$):} \\
The resulting neutrino mass in this case is consistent
with oscillation data, $m_\nu \sim {\rm eV}$, for a NP scale
of several TeV and ${\cal O}(1)$
couplings, i.e., $f_{s^2 \phi^2} \cdot \left( y_{s \nu}^L \right)^2 \sim {\cal O}(1)$.
Therefore, here also,
no useful bound can be put on the corresponding dimension 5 operator
$\bar s s^c \phi^2$.
\item{b-quark case ($y_{q \ell}^L = y_{b \nu}^L$ and $f=f_{b^2 \phi^2}$):} \\
This corresponds to the 3rd generation LQ case, for which we obtain
$m_\nu \sim {\rm KeV}$ with $f_{b^2 \phi^2} \cdot \left( y_{b \nu}^L \right)^2
\sim {\cal O}(1)$ and a NP scale
of several TeV. Thus, in this case, the neutrino mass bound
constrains the dimension 5 operator $\bar b b^c \phi^2$
or the corresponding LQ couplings:
either $\Lambda_{b^2 \phi^2} \sim {\cal O}(1000)$ TeV or
$f_{b^2 \phi^2} \cdot \left( y_{b \nu}^L \right)^2
\sim {\cal O}(10^{-3})$.
\end{description}

Finally, we wish to further comment on the link between neutrino masses and the underlying heavy physics.
As noted in the previous section, the heavy fermionic states $\Psi(1,1,0)$ and $\Psi(1,3,0)$ can generate at tree-level
both the Weinberg operator
$\bar\ell^c \tilde H^\star  \tilde H^\dagger \ell$ and the operator
$\bar\ell d \tilde H \phi^*$, while
$\Psi(1,1,0)$ can generate all three types of dimension
five operators $\bar\ell^c \tilde H^\star  \tilde H^\dagger \ell$, $\bar\ell d \tilde H \phi^*$ and
$\bar d d^c \phi^2$.
Therefore, in this setup there are several scenarios that do not require small coupling constants:
\begin{enumerate}
\item The heavy fermionic state $\Psi(1,1,0)$ is responsible
for generating all dimension five operators $\bar\ell^c \tilde H^\star  \tilde H^\dagger \ell$, $\bar\ell d \tilde H \phi^*$ and
$\bar d d^c \phi^2$, with a typical mass scale of $M_\Psi \sim {\cal O}(10^{14})$ GeV.
In this case, the Majorana neutrino mass term will be generated at tree-level through the
type I seesaw mechanisms by the Weinberg operator $\bar\ell^c \tilde H^\star  \tilde H^\dagger \ell$,
whereas the 1-loop and 2-loops contribution from the operators
$\bar\ell d \tilde H \phi^*$ and $\bar d d^c \phi^2$ will be negligible.
\item The heavy fermionic state $\Psi(1,3,0)$ is responsible
for generating both operators $\bar\ell^c \tilde H^\star  \tilde H^\dagger \ell$ and
$\bar\ell d \tilde H \phi^*$, with a typical mass scale of $M_\Psi \sim {\cal O}(10^{14})$ GeV,
while the operator $\bar d d^c \phi^2$ is generated by another heavy mediator.
In this case, the Majorana neutrino mass term can be generated again at tree-level
through the type I or type III seesaw
mechanisms by the Weinberg operator $\bar\ell^c \tilde H^\star  \tilde H^\dagger \ell$
and the 1-loop contribution from the operator $\bar\ell d \tilde H \phi^*$ will be subdominant.
This holds also in the case that the Weinberg operator is generated by the heavy scalar
$\Phi(1,3,0)$ if $M_\Phi \sim {\cal O}(10^{14})$ GeV and a corresponding
${\cal O}(1)$ Wilson coefficient. Note that, in this case, a 2-loop
Majorana mass term can be generated as well by the operator $\bar d d^c \phi^2$,
depending on the couplings involved (see discussion above).
\item The Weinberg operator is not relevant to neutrino masses, i.e.,
there are no heavy $\Phi(1,3,0)$, $\Psi(1,1,0)$ and $\Psi(1,3,0)$ states in the
underlying theory. In this case, neutrino  masses are not generated through the
seesaw  mechanism, but they may be still generated at 1-loop or at 2-loops by
the dimension five operators $\bar\ell d \tilde H \phi^*$ and
$\bar d d^c \phi^2$ as described above, if these operators
are generated at
tree-level in the underlying theory by other heavy states (see previous section).
\end{enumerate}

\subsection{Neutrinoless double beta decay \label{sec43}}

\begin{figure}[htb]
\begin{center}
\includegraphics[scale=0.5]{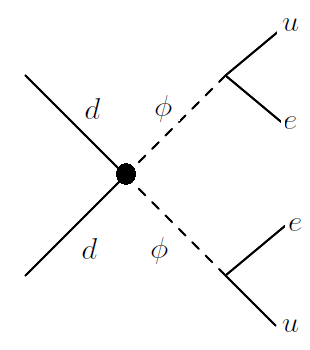}
\end{center}
\vspace{-0.8cm}
\caption{Tree-level graph that generates neutrinoless double beta decay
 via the dimension five operator $\bar d d^c \phi^2$. See also text.}
\label{2beta}
\end{figure}

The dimension five operator $\bar d d^c \phi^2$ can mediate neutrinoless double beta decay ($0\nu\beta\beta$)
via the diagram depicted in Fig.~\ref{2beta}. This requires
both the dimension five operator
$\bar d d^c \phi^2$ and the Yukawa-like renormalizable coupling
of $\phi$ to the right-handed 1st generation u-quark and
electron, i.e., the term  $\propto y_{u e}^R$
in ${\cal L}_{Y,\phi}$ (see Eq.~\ref{phiSMY}). If $\phi$ is a 3rd generation
leptoquark,  we expect $y_{u e}^R \ll 1$ (see discussion
in the next section) in which case the $0\nu\beta\beta$ decay rate will be significantly suppressed.

The limit on $0\nu\beta\beta$ decay is usually expressed in terms of the electron-electron
element of the neutrino mass matrix. The current
 bound is $|(m_\nu)_{ee}| < 0.1 -0.5$ eV, depending
 on the $0\nu\beta\beta$ experiment, see e.g., \cite{0nubeta}.
This translates into a bound on the corresponding parton-level amplitude for
$0\nu\beta\beta$ \cite{jose1}:
\begin{eqnarray}
\frac{p_{eff}}{G_F^2} |{\cal A}_{0\nu\beta\beta}| \simeq \frac{|(m_\nu)_{ee}|}{p_{\rm eff}}
< 5 \times 10^{-9}  \label{bound1} ~,
\end{eqnarray}
where $p_{\rm eff} \sim 100$ MeV is the neutrino effective momentum obtained by averaging the
corresponding nuclear matrix element contribution.

In our case, the $0\nu\beta\beta$ amplitude corresponding to the diagram in Fig.~\ref{2beta}
can be estimated as:
\begin{eqnarray}
{\cal A}_{0\nu\beta\beta} \sim \frac{f \cdot |y_{u e}^R|^2}{\Lambda M_\phi^4} ~.
\end{eqnarray}
where $f=f_{d^2 \phi^2}$ and $\Lambda = \Lambda_{d^2 \phi^2}$. Therefore, using Eq.~\ref{bound1} we obtain:
\begin{eqnarray}
\frac{\Lambda}{\rm TeV} \gsim  150 \cdot \frac{f \cdot |y_{u e}^R|^2}{(M_\phi/{\rm TeV})^4} ~.
\end{eqnarray}

In particular, we find that no useful bound can be imposed on
the scale
of the dimension five operator $\bar d d^c \phi^2$,
assuming $f_{d^2 \phi^2} \sim {\cal O}(1)$ and a TeV-scale
LQ mass, $M_{\phi} \sim {\cal O}(1~ {\rm TeV})$, if the LQ $\phi$ is a 3rd generation LQ (as assumed below), i.e.,
having a suppressed Yukawa-like coupling to the 1st generation right-handed fermions: $y_{u e}^R < 0.1$.

\medskip

\section{Collider phenomenology of a 3rd generation scalar Leptoquark in the EFT \label{sec5}}

We next discuss the expected NP signals of the down-type  $\phi(3,1,-\frac{1}{3})$
and up-type   $\phi(3,1,\frac{2}{3})$ LQs
at the 13 TeV LHC and also at future higher energy hadron colliders such as
a 27 TeV High-Energy LHC (HE-LHC) and a 100 TeV Future Circular proton-proton
Collider (FCC-hh) \cite{fcc}.

All cross-sections presented
in this section were calculated
using MadGraph5 \cite{madgraph5}
at LO parton-level,
for which  a dedicated universal FeynRules output (UFO) model
for the LQ-SM EFT framework defined in Eq.~\ref{EFT1}
was produced for the MadGraph5 sessions
using FeynRules \cite{FRpaper}.
The LO nnpdf3 PDF set (NNPDF30-lo-as-0130 \cite{PDFset})
was used in all the calculations presented below.
Also, all cross-sections were calculated with a dynamical scale choice for the
central value of the factorization ($\mu_F$)
and renormalization ($\mu_R$) scales corresponding to the sum of
the transverse mass in the hard-process, and, for consistency with the EFT
framework, a cut on the center of mass energy
of $\sqrt{\hat s} < \Lambda$ was placed using Mad-Analysis5 \cite{madanal},
where several values of $\Lambda$
(the scale of NP) were used for the processes considered
below.$^{4}$

Furthermore, we will assume throughout the rest of the paper
that $\phi(3,1,-\frac{1}{3})$ and $\phi(3,1,\frac{2}{3})$, under
consideration in this section,
are 3rd generation leptoquarks and denote them generically by $\phi_3$.
In particular, we assume that the LQ-lepton-quark Yukawa-like
couplings of $\phi_3$ to the 1st and 2nd
generations SM fermions in the corresponding renormalizable $\phi$SM Lagrangian
are much smaller than its couplings to the 3rd generation quark-lepton pair,
e.g., to a $t \tau$ and/or $b \nu_\tau$ pairs in the case of the down-type LQ
$\phi(3,1,-\frac{1}{3})$ (see Eqs.~\ref{phiSMY}).

This scenario can be realized by imposing an approximate
$Z_3$ generation symmetry under
which the physical states of the SM fermions (i.e., mass eigenstates)
transform as:
\begin{eqnarray}
\psi^k \to e^{i \alpha({\psi^k}) \tau_3} \psi^k ~,
~~ \tau_3 \equiv 2\pi/3 \label{Z3} ~,
\end{eqnarray}
where $k$ is the generation index and $\alpha({\psi^k})$
are the $Z_3$ charges of $\psi^k$.

Consider for example the down-type LQ $\phi(3,1,-\frac{1}{3})$:
if the $Z_3$ charges equal the generation index,
i.e., $\alpha(\psi^{k}) = k$, and
$\alpha(\phi) = 3 $, then only terms in ${\cal L}_{\phi SM}$
involving the 3rd generation are allowed. In particular, assuming
Baryon number conservation and thus ignoring
the $Z_3$-allowed LQ interactions
with the 3rd generation quarks (i.e., $\phi \bar t^c_R b_R$
and $\phi \bar t_L^c b_L$) that would in general allow for
proton decay, we have:
\begin{eqnarray}
{\cal L}_{Y,\phi_3} \approx y_{q_3\ell_3}^L \left( \bar t_L^c \tau_L +
\bar b_L^c \nu_{\tau L} \right) \phi^* +
y_{u_3e_3}^R \bar t^c_R \tau_R \phi^*
+ \mbox{H.c.} \label{phi3SMY} ~.
\end{eqnarray}
where we will assume that
the above Yukawa-like LQ-quark-lepton 3rd generation couplings
are ${\cal O}(1)$.

The $Z_3$ generation symmetry is exact in the limit where
the quark mixing CKM matrix $V$ is diagonal, so that $Z_3$-breaking effects
will in general be proportional
to the square of the small off-diagonal CKM elements $|V_{cb}|^2$,
$|V_{ub}|^2$, $|V_{ts}|^2$, $|V_{td}|^2$, and will, therefore, be suppressed (see also \cite{bparity,1810.11588,1808.00942}).
In particular, the $Z_3$ generation symmetry is assumed to be broken in the underlying heavy theory and can, therefore, be traced to the higher dimensional operators.
For example, the off-diagonal SM Yukawa couplings may be generated by
the dimension six operators:
\begin{eqnarray}
\Delta {\cal L}_{Y,H}^{(6)} =
\left( f_{uH} \bar q_L \tilde H u_R + f_{dH} \bar q_L H d_R \right) \frac{H^{\dagger} H}{\Lambda^2} + \mbox{H.c.}
\label{eq1}\,,
\end{eqnarray}
where, if e.g., $\Lambda \sim 1.5,3$ or $5$ TeV and $f_{uH},f_{dH} \sim {\cal O}(1)$,
then the resulting effective Yukawa couplings,
$y_{\rm eff} = f_{uH,dH} \cdot v^2/\Lambda^2$,
are $y_{\rm eff} \sim {\cal O}(y_b^{SM})$,
$y_{\rm eff} \sim {\cal O}(y_c^{SM})$
or $y_{\rm eff} \sim {\cal O}(y_s^{SM})$,
respectively, where $y_q^{SM}$ are the corresponding Yukawa couplings
in the SM (see \cite{universalY}).

The $Z_3$  breaking terms in the LQ sector
will also be generated in the effective theory through  higher dimensional
operators. To demonstrate that consider for example the
dimension five operator
$\bar d d^c \phi^2$ in Eq.~\ref{d2p2dim5}. As was shown in section \ref{sec3},
this operator can be generated
at tree-level in the UV theory by exchanging e.g., a heavy scalar $\Phi(6,1,-\frac{2}{3})$
(see diagram (a) in Fig.~\ref{figd2p2}).
Thus, if $\Phi(6,1,-\frac{2}{3})$ couples to
the 1st and/or 2nd generation down-quarks, then the $Z_3$ generation symmetry is broken and
the scale of generation breaking is the mass of $\Phi(6,1,-\frac{2}{3})$, $M_{\Phi}$.
In particular, the $Z_3$ generation breaking effects in this case will be proportional to
$g_{\Phi dd} \cdot g_{\Phi \phi \phi} / M_{\Phi}$,
where $g_{\Phi dd}$ and $g_{\Phi \phi \phi}$ are the
couplings
of the heavy $\Phi(6,1,-\frac{2}{3})$ to a $dd$-pair and a
$\phi \phi$-pair, respectively. The matching
to the EFT framework
of Eq.~\ref{dim5all} can be done by replacing
$M_{\Phi} \to \Lambda_{d^2 \phi^2}$ and
$g_{\Phi dd} \cdot g_{\Phi \phi \phi} \to f_{d^2 \phi^2}$.
%
%

We thus, allow for higher dimensional interactions of $\phi_3$ with the lighter
SM fermion generations, keeping in mind that these are a-priori suppressed in the EFT by
inverse powers of the NP scale (e.g., by $1/\Lambda$
if it originates from the dimension five operators) and that,
in this case, $\Lambda$ represents the scale of breaking
the $Z_3$ generation symmetry.$^{5}$
\footnotetext[4]{The UFO model files are available upon request.}
\footnotetext[5]{Note that the couplings of
$\phi_3$ to the 1st and 2nd generations fermions can also
be loop generated by the renormalizable LQ-quark-lepton couplings. In this case they are suppressed by the corresponding loop factor and CKM elements and are, therefore, subdominant.}

\subsection{The down-type scalar LQ $\phi(3,1,-\frac{1}{3})$}

We now consider the LHC signals of
the down-type 3rd generation LQ $\phi_3=\phi_3(3,1,-\frac{1}{3})$ under investigation.
Following our above setup where $\phi_3$ is expected to have suppressed couplings to 1st and 2nd generation fermions,
single $\phi_3$ production will occur through the channel $gb \to \phi_3 \nu_\tau$,
with a cross-section
 $\sigma(p p_{(gb)} \to \phi_3 \nu_\tau) \sim 3.5(0.025)$ fb for $M_{\phi_3} = 1(2)$ TeV and
$y_{b \nu_\tau}^L=1$ \cite{1801.07641}.
Also, with sub-leading couplings to the 1st and 2nd generation fermions,
the main channels for
$\phi_3$ pair-production will be gluon
and $q-\bar q$ fusion,
where the typical cross-sections are
$\sigma(p p_{(gg,q \bar q)} \to \phi_3 \phi_3^*) \sim 5.5(0.01)$ fb for $M_{\phi_3} = 1(2)$
TeV \cite{1801.07641} (with no cut on the $\phi_3 \phi_3^*$ invariant mass) and do not depend on
the $\phi_3$-quark-lepton
couplings.
Thus, assuming that $\phi_3$ decays via $\phi_3 \to t \tau^-$ and/or $\phi_3 \to b \nu_\tau$ with
50\% branching ratio into each channel, we find e.g.,
$\sigma(p p_{(gg,q \bar q)} \to \phi_3 \phi_3^* \to t \bar t \tau^- \tau^+) \sim 1.4$ fb
at a 13 TeV
LHC if $M_{\phi_3} \sim 1$ TeV.
A dedicated search in this channel was carried by CMS in \cite{1803.02864}, where
no evidence for this signal was found, setting a limit on the
LQ mass of $M_{\phi_3} \gsim 900$ GeV at 95\% confidence level for $BR(\phi_3 \to t \tau^-) =1$.

As mentioned above, LQ phenomenology changes in the presence of the higher dimensional
effective operators. In particular, additional potentially interesting
$\phi_3$ production channels are opened at the LHC. However,
most of them will have a too small cross-section
at the 13 TeV LHC, due to the $1/\Lambda^n$ suppression in the EFT expansion,
so that the leading effects are produced by the dimension five
operators involving $\phi_3$ in Eq.~\ref{dim5all}. Recall, however, that the operator
$\bar\ell d \tilde H \phi^*$ is expected to have suppressed
effects because of a large effective scale,
as required for consistency with sub-eV neutrino masses (cf. the previous section).

\begin{figure}[htb]
\begin{center}
\includegraphics[scale=0.22]{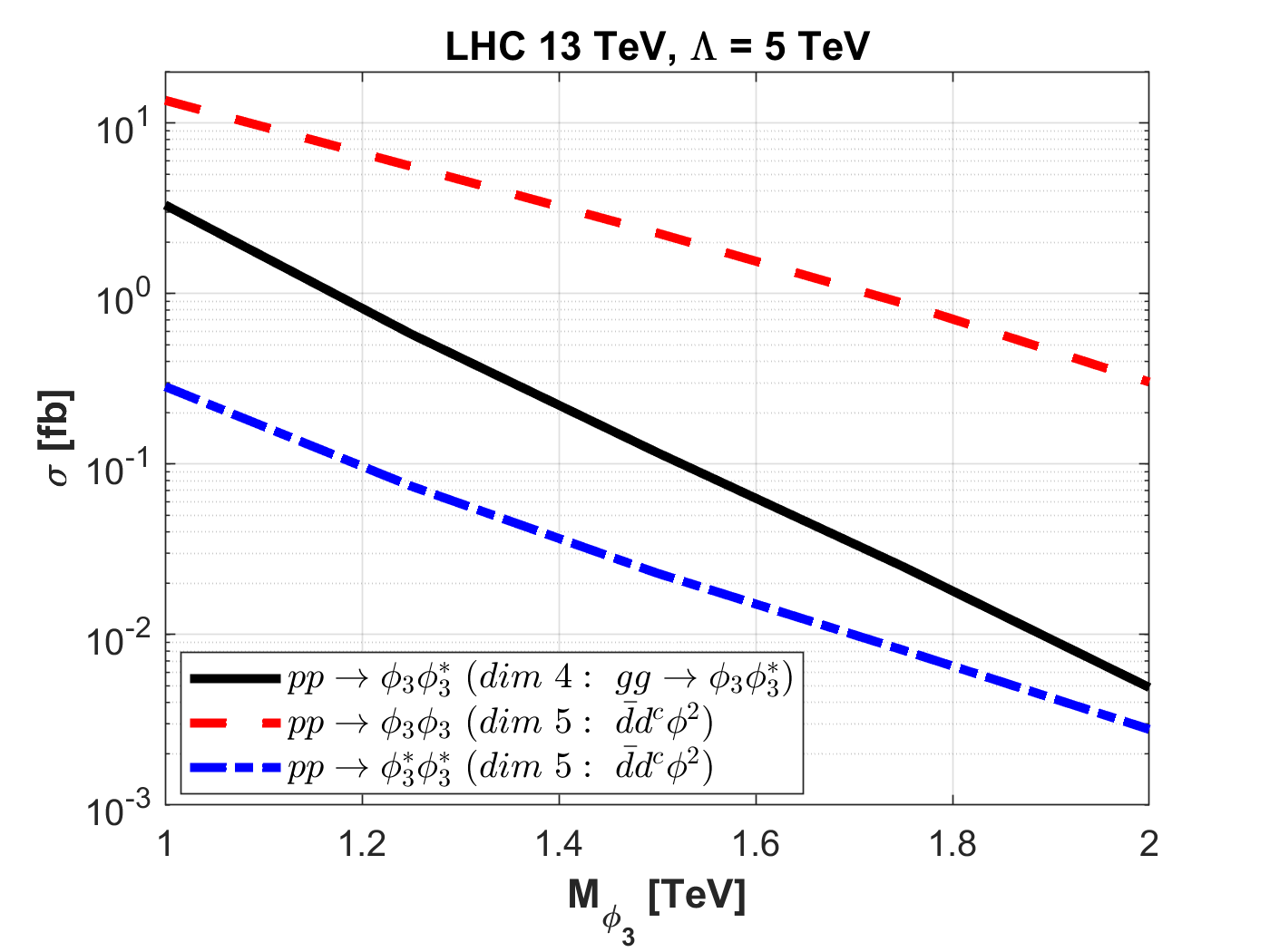}
\end{center}
\vspace{-0.8cm}
\caption{Pair-production cross-sections of the down-type LQ  $\phi_3$ at the 13 TeV LHC
with $\Lambda_{d^2 \phi^2} = 5$ TeV:
$pp \to \phi_3 \phi_3$ (dashed line), $pp \to \phi_3 \phi_3^*$ (solid line) and
$pp \to \phi_3^* \phi_3^*$ (dashed-dot line) (see also text).}
\label{LQCSX}
\end{figure}

We are therefore left with only one dimension five operator,
$\bar d d^c \phi^2$, that can potentially mediate interesting $\phi_3$ pair-production
signals at the LHC. In particular, we find that this operator may yield a strikingly large {\it asymmetric} same-sign(charge) $\phi_3 \phi_3$ signal at the LHC via
$dd \to \phi_3 \phi_3$, which is more than an order of magnitude larger
than the charged conjugate channel
$\bar d \bar d \to \phi_3^* \phi_3^*$, due to the different fractions
of $d$ and $\bar d$ in the incoming protons, see Fig.~\ref{LQCSX}.
The hard cross-section for $dd \to \phi_3 \phi_3$ (which equals that of
the charged conjugate one $\bar d \bar d \to \phi_3^* \phi_3^*$) is:
\begin{eqnarray}
\hat\sigma(dd \to \phi_3 \phi_3) = \frac{\beta f^2}{12 \pi \Lambda^2} ~,
\end{eqnarray}
where (cf. Eq.~\ref{dim5all}) $\Lambda = \Lambda_{d^2 \phi^2}$,
 $f =f_{d^2 \phi^2}$,
 $\beta^2 =  1- 4M_{\phi_3}^2/\hat s$, and $\sqrt{\hat s}$ is the center of mass
energy of the hard process. For example, if
$\Lambda_{d^2 \phi^2} = 5$ TeV (and with a cut
on the $\phi_3 \phi_3$ invariant mass, $M_{\phi_3 \phi_3} < 5$ TeV),
we find:$^{6}$\footnotetext[6]{There are no SM contributions to the processes studied
here and also none of the
tree-level generated dimension six operators that we
list in the appendix contribute to them.
Furthermore, other dimension six operators which do not involve the LQ fields
and which can, in principle, be generated by the heavy mediators in Figs.~\ref{figd2p2}
and \ref{figldhp} (e.g., four-fermion operators such as $d \bar d \ell^+ \ell^-$
and $(d \bar d)^2$), do not affect the same-sign lepton signals considered in this work.
Thus, the dimension five operators that we consider generate the leading contributions
to these processes. In particular,
potential corrections to the leading-order cross-sections presented
in this section can be generated either by loop-generated dimension six operators
and/or by dimension seven operators.
The former are
suppressed by a factor of $E/(16 \pi^2 \Lambda)$
($E$ is the typical energy of the process) and can, therefore, be neglected here, while
the latter are suppressed typically by $(E/\Lambda)^2$
and, therefore, their size depend
on the relevant energy scale of the process.
In particular, for the s-channel process (see Fig.~\ref{figd2p2}a) the corrections can reach $~50\%$,
while for t or u channel processes (see Fig.~\ref{figd2p2}b) the relevant energy scale is much
smaller and the corrections are again negligible.}
\begin{eqnarray}
\sigma(pp \to \phi_3 \phi_3)_{M_{\phi_3} \sim 1~{\rm TeV}} &\sim& 14 ~{\rm fb} ~, \nonumber \\
\sigma(pp \to \phi_3 \phi_3)_{M_{\phi_3} \sim 2~{\rm TeV}} &\sim& 0.3 ~{\rm fb} \label{signals1}~.
\end{eqnarray}

This can be compared to the
gluon-fusion cross-section of the opposite-charge $\phi_3 \phi_3^*$
pair-production signal, $pp_{(gg)} \to \phi_3 \phi_3^*$,
for which the hard cross-section (see e.g., \cite{9610408,9704322}):
%
\begin{eqnarray}
\hat\sigma(gg \to \phi_3 \phi_3^*) &=& \frac{\pi \alpha_s^2}{96 \hat s} \cdot
\left\{ \beta (41 -31\beta^2) \right.  \\
\hspace{-0.5cm}&-& \left. (17-18 \beta^2 + \beta^4) \cdot
log \left( \frac{1+\beta}{1-\beta} \right) \right\}  ~, \nonumber
\end{eqnarray}
%
drops with the energy as $1/\hat s$ and yields a cross-section of
(again with $M_{\phi_3 \phi_3^*} < 5$ TeV):
\begin{eqnarray}
\sigma(pp \to \phi_3 \phi_3^*)_{M_{\phi_3} \sim 1~{\rm TeV}} &\sim& 3 ~{\rm fb} ~, \nonumber \\
\sigma(pp \to \phi_3 \phi_3^*)_{M_{\phi_3} \sim 2~{\rm TeV}} &\sim& 0.005 ~{\rm fb} \label{signals2}~.
\end{eqnarray}

We thus see that the same-sign $\phi_3 \phi_3$ rate is expected to be larger than
the opposite-sign $\phi_3 \phi_3^*$ rate at the 13 TeV LHC, in particular,
$\sigma(pp \to \phi_3 \phi_3)/\sigma(pp \to \phi_3 \phi_3^*) \sim 5(60)$ for $M_{\phi_3}=1(2)$ TeV.

Taking into account the leading $\phi_3$ decays $\phi_3 \to t \tau^-$ and $\phi_3 \to b \nu_\tau$,
this signal will in turn give rise to
the new asymmetric signatures ($j_b = b$-jet):
\begin{itemize}
\item $pp \to \phi_3 \phi_3 \to 2 \cdot j_b + \missET$
\item $pp \to \phi_3 \phi_3 \to tt \tau^- \tau^-$
\item $pp \to \phi_3 \phi_3 \to t \tau^- + j_b + \missET$
\end{itemize}
with a cross-section which is more than an order of magnitude larger than
the charged conjugate channels.

While $pp \to  2 \cdot j_b + \missET$ may not be unique to $\phi_3$ pair-production,
and may be more challenging due to the larger background expected in this channel,
the signal of same-sign
top-quark pair in
association with a pair of same-sign negatively charged $\tau$-leptons, $pp \to tt \tau^- \tau^-$,
and the single top - single $\tau$ signature, $pp \to t \tau^- + j_b + \missET$,
may give striking new asymmetric $\phi_3 \phi_3$ signals.

For example, if the scale
of the NP underlying ${\cal L}_{\phi SM}$ is $\Lambda =5$ TeV,
the LQ mass is $M_{\phi_3} \sim 1$ TeV and its leading
branching ratios are $BR(\phi_3 \to t \tau^-)=BR(\phi_3 \to b \nu_\tau)=0.5$, then
we expect
$\sigma(pp \to tt \tau^- \tau^-) \sim 3.4$ fb;
while $\sigma(pp \to \bar t \bar t \tau^+ \tau^+) \sim 0.07$ fb,
see Fig.~\ref{LQCSX}. The former is
about five times larger than the rate for the gluon-fusion
$\phi_3 \phi_3^*$ signal $pp \to t \bar t \tau^+ \tau^-$, for which a dedicated search has already been performed by
CMS \cite{1803.02864} with null results.

\footnotetext[7]{This estimate does not include the $\tau$-decay branching ratio into a specific final state.}
With an integrated luminosity of $\sim$300  fb$^{-1}$,
$\Lambda =5$ TeV and $M_{\phi_3} \sim 1$ TeV, about 1000
$tt \tau^- \tau^-$ events with an invariant
mass smaller than 5 TeV are expected.
After the top-quarks decay hadronically via
$t \to W^+ b \to 2 \cdot j+ b$ ($j=$light jet) with a
$BR(t \to W^+ b \to 2 \cdot j+ b) \sim 2/3$,
we expect about 450 same-sign $\tau^- \tau^-$ events with a high jet-multiplicity
signature: $pp \to \tau^- \tau^- + 4 \cdot j + 2 \cdot j_b$ and with
a statistical error of $\sim \sqrt{450} \sim 20$ events and no irreducible
background (see also discussion below).$^{7}$
Note also that roughly the same number of events are expected for the $t \tau^-$ production signal
$pp \to t \tau^- + j_b + \missET$, which leads to $pp \to \tau^- + 2 \cdot j + 2 \cdot j_b + \missET$,
when the top-quark decays hadronically via $t \to W^+ b \to 2 \cdot j+ b$. This
single-$\tau$ signal lack a unique characterization akin the same-sign
lepton signature in pair LQ production and might, therefore, be harder to trace.

It is also useful to define the inclusive same-charge $\tau \tau$ asymmetry:
%
\begin{eqnarray}
{\cal A}_{\tau \tau} \equiv \frac{\sigma(pp \to \tau^- \tau^- + X_{j}) - \sigma(pp \to \tau^+ \tau^+ + X_{j})}{\sigma(pp \to \tau^- \tau^- + X_{j}) + \sigma(pp \to \tau^+ \tau^+ + X_{j})} \label{asym} ~,
\end{eqnarray}
%
where we have assumed again that the top-quark decays hadronically
via $t \to W^+ b \to 2 \cdot j+ b$ and $X_{j}$ stands for any accompanying jets
in the final state, i.e., for events with prompt same-sign $\tau \tau$ and
no missing transverse energy (MET). When the $\phi_3$ mass is in the range
$1~{\rm TeV} \lsim M_{\phi_3} \lsim 2~{\rm TeV}$,
we expect ${\cal A}_{\tau \tau} \to 1$
since this
asymmetry receives its most significant contribution from the $\phi_3 \phi_3$ and $\phi_3^* \phi_3^*$
channels (see Fig.~\ref{LQCSX}).
The SM background for the same-sign
$\tau^- \tau^-$ events with no MET, from processes
that can mimic this final state, is expected to be significantly suppressed,
in particular, after imposing the appropriate kinematical and selection cuts
(see also comment below).
We, therefore, expect the above same-charge asymmetry
${\cal A}_{\tau \tau}$ to be close to a 100\%.

The statistical significance, $N_{SD}$, with which this asymmetry can
be detected at the LHC is:
\begin{eqnarray}
N_{SD} \sim \sqrt{\sigma_{\tau \tau} \cdot L} \cdot {\cal A}_{\tau \tau} \cdot \sqrt{\epsilon}
\label{NSD}~,
\end{eqnarray}
where $\sigma_{\tau \tau}$ is the inclusive cross-section
$\sigma(pp \to \tau^- \tau^- + X_{j})$
and $\epsilon$
is the corresponding combined efficiency for the simultaneous measurement
of this final state.
Thus, with an integrated luminosity of 300 inverse fb (recall
that $\sigma(pp \to \tau^- \tau^- + X_{j}) \sim 1.5$ fb for $\Lambda = 5$ TeV and
$M_{\phi_3} = 1$ TeV), and
a combined efficiency of $\epsilon \sim 0.01$, this asymmetry
can be detected with about a $\sim 2\sigma$ significance. At the
high-luminosity LHC with 3000 inverse fb this asymmetry should be accessible with a statistical significance of $N_{SD} \sim 7$.

Finally, we wish to further comment on the potential background
to the LNV same-sign lepton signals considered
here and in the following section. Although these signals have formally
no irreducible SM background (since lepton number is conserved in the SM), they can be ''contaminated" by reducible background that
can mimic these signatures due to higher-order effects (e.g., initial
and final state radiation), particle/jets
miss-identification, $\tau^\pm$ reconstruction limitations, heavy flavor decays and alike.
However, due to the distinct characteristics of our
same-sign (isolated) lepton-pair signals,
such a background can
in principle be reduced to the desired level with the appropriate kinematical and
selection cuts as well as veto requirements, e.g.,  on the MET and the energy
distribution of the jets in the process, see e.g., the recent SUSY searches
in same-sign lepton events at the LHC,
performed by the CMS \cite{SSCMS} and ATLAS \cite{SSATLAS}
collaborations.
An example of such a potential background is the SM process
$pp \to t(\to b W^+ \to b jj)
\bar t(\to \bar b W^- \to \bar b \tau^- \bar\nu_\tau) W^-(\to \tau^- \nu_\tau)$ and
the charged conjugate channel (considered also in \cite{SSCMS,SSATLAS}), which
lead to same-sign $\tau^\pm \tau^\pm$ events
that can mimic our LNV LQ mediated signals, e.g., from
$pp \to t(\to b W^+ \to b jj) t(\to b W^+ \to b jj)
\tau^- \tau^-$. This SM process can, therefore, also
''contaminate" the asymmetry ${\cal A}_{\tau \tau}$ in Eq.~\ref{asym}, since
$\sigma(pp\to t \bar t W^+) \sim 2 \sigma(pp \to t \bar t W^-)$.
However, not only that this background has a cross-section
of the same order of the LNV signal
considered above, i.e.,
$\sigma(pp \to t \bar t W^\pm \to \tau^\pm \tau^\pm + X_j + \missET) \sim
\sigma(pp \to t t \tau^- \tau^- \to \tau^- \tau^- + X_j) \sim {\cal O}(1)$ fb,
it also contains a different energy distribution of the MET and jets
in the process and can, therefore, be significantly reduced with the
proper selection cuts and veto requirements.

\subsection{The up-type scalar LQ $\phi(3,1,\frac{2}{3})$}

We wish to briefly comment here on the phenomenology and LHC
signals expected for an up-type LQ $\phi(3,1,\frac{2}{3})$ in the EFT framework.
The renormalizable Yukawa-like interactions of this LQ contain
only the term $y_{d^i d^j}^R \bar d_{R}^{ci} d_R^j \phi$, where
$y_{d^i d^j}^R$is anti-symmetric due to SU(3) (color) gauge invariance.
Note, however, that this di-quark LQ coupling violates baryon number and, in the presence of the higher dimensional LQ couplings to
quark-lepton pairs (see below), may mediate proton decay.
We therefore, assume that it is either negligibly small or forbidden
due to a symmetry.

\footnotetext[8]{We note that
if both the down-type and up-type LQ are included as light degrees of freedom
in the low-energy framework, then four more dimension five operators can be
constructed in the EFT extension:
$\bar q \ell^c \phi_d^* \phi_u^*$,
$\bar u e^c \phi_d^* \phi_u^*$,
$\bar q q^c \phi_d \phi_u$ and
$\bar d u^c \phi_d \phi_u$, where we have used here
the subscripts $d$ and $u$ to distinguish between them.}

In the up-type $\phi(3,1,\frac{2}{3})$ case,
we find that there are
four dimension five operators (in addition to the Weinberg operator of Eq.~\ref{h2l2dim5}):$^{8}$
\begin{widetext}
\begin{eqnarray}
\Delta {\cal L}_{\phi SM}^{(5)} =
\frac{f_{\ell u \phi H}}{\Lambda_{\ell u \phi H}} \bar\ell u \tilde H \phi^* +
\frac{f_{\ell d \phi H}}{\Lambda_{\ell d \phi H}} \bar\ell d H \phi^* +
\frac{f_{q e \phi H}}{\Lambda_{q e \phi H}} \bar q e H \phi +
\frac{f_{u^2 \phi^2}}{\Lambda_{u^2 \phi^2}} \bar u u^c \phi^2 +\mbox{ H.c.} \label{dim5up}~.
\end{eqnarray}
\end{widetext}

The fourth operator in Eq.~\ref{dim5up},
$\bar u u^c \phi^2$, will
give rise to a similar
same-sign asymmetric $\phi_3 \phi_3$ signals via
$uu \to \phi_3 \phi_3$ (and the much smaller
charged conjugate one $\bar u \bar u \to \phi_3^* \phi_3^*$), with a considerably larger cross-section
than the same-sign down-type LQ pair-production one,
due to the larger $u$-quark
content/PDF in the protons.
For example, with $\Lambda_{u^2 \phi^2} = 5$ TeV and the invariant
mass cut $M_{\phi_3 \phi_3} < 5$ TeV,
we find for the up-type LQ case:
\begin{eqnarray}
\sigma(pp \to \phi_3 \phi_3)_{M_{\phi_3} \sim 1~{\rm TeV}} &\sim& 77 ~{\rm fb} ~, \nonumber \\
\sigma(pp \to \phi_3 \phi_3)_{M_{\phi_3} \sim 2~{\rm TeV}} &\sim& 3 ~{\rm fb} \label{signals3}~, \nonumber \\
\end{eqnarray}
which is about 25(600) times larger than the expected opposite-charged $\phi_3 \phi_3^*$ signal for
$M_{\phi_3}=1(2)$ TeV, see Eq.~\ref{signals2}.

In contrast  to the case of the down-type LQ (which decays
via its renormalizable couplings to quark-lepton pairs),
the decay pattern of the up-type LQ considered here will be controlled by
its dimension five interactions with the SM fields in Eq.~\ref{dim5up}.
In particular, it will decay via either $\phi \to d e^+$
and/or $\phi \to u \nu$, where $d,u,e,\nu$ stand here for
a down-quark, up-quark, charged lepton and neutrino of any generation,
with a corresponding coupling which is
suppressed by $\sim v/{\Lambda}$, e.g., for the decay $\phi \to u \nu$
the coupling is $f_{l u \phi H} \cdot (v/\Lambda_{l u \phi H})$.
Thus, assuming as an example that its dominant
dimension five couplings are to the 3rd generation SM fermions, then
here also,
when it decays via
either $\phi_3 \to b \tau^+$ and/or $\phi_3 \to t \nu_\tau$, we
expect the new asymmetric signals:
\begin{itemize}
\item $pp \to \phi_3 \phi_3 \to tt + \missET$
\item $pp \to \phi_3 \phi_3 \to \tau^+ \tau^+ + 2 \cdot j_b$
\item $pp \to \phi_3 \phi_3 \to t \tau^+ + j_b + \missET$
\end{itemize}
each having a cross-section which is
several orders of magnitude larger than
the charged conjugate channels.

Despite obvious parallels, there are important differences between the above signals and the ones expected
for the down-type LQ :
\begin{enumerate}
\item The same-sign $\tau^+ \tau^+$ signal $pp \to \phi_3 \phi_3 \to \tau^+ \tau^+ + 2 \cdot j_b$ for the up-type LQ
has opposite lepton charges than the corresponding signal for the down-type LQ.
Therefore, the asymmetry ${\cal A}_{\tau \tau}$ flips signs in the up-type LQ case.
\item Similarly, in single LQ production, the final $\tau$ lepton is positive for the up-type LQ and negative for the down-type.
\item The same-sign $\tau \tau$ signal has a lower jet multiplicity than in the case of the down-type LQ.
\item The same-charge top-quark pair signal $pp \to \phi_3 \phi_3 \to tt + \missET$
can also yield a same-sign lepton signal $ pp \to \ell^+ \ell^+ + 2 \cdot j_b + \missET$,
involving any of the charged leptons, i.e., $\ell^+ \ell^+ = e^+e^+, \mu^+ \mu^+, \tau^+ \tau^+$,
if the top-quark decays leptonically via $t \to W^+ b \to \ell^+ \nu_\ell b$.
\end{enumerate}

Thus, the most promising signals in up-type
$\phi_3 \phi_3$ pair-production
are $pp \to \tau^+ \tau^+ + 2 \cdot j_b$ and
$pp \to tt + \missET \to \ell^+ \ell^+ + 2 \cdot j_b + \missET$,
containing two positive charged leptons (for which the background is low) and
two high-$p_T$ tagged b-jets.
For $\Lambda=5$ TeV, $M_{\phi_3} =1$ TeV and assuming
$BR(\phi_3 \to b \tau^+)=BR(\phi_3 \to t \nu_\tau)=0.5$,
the overall cross-sections for these signals (with an invariant mass
smaller than 5 TeV) are expected to be:
\begin{eqnarray}
&& \sigma(pp \to \tau^+ \tau^+ + 2 \cdot j_b) \sim 20 ~ {\rm fb} ~, \nonumber \\
&& \sigma(pp \to \ell^+ \ell^+ + 2 \cdot j_b + \missET) \sim 0.2 ~ {\rm fb} ~,
\end{eqnarray}
where, as mentioned above,
for the same-charged top-quark pair signal,
$pp \to tt + \missET \to \ell^+ \ell^+ + 2 \cdot j_b + \missET$,
this cross-section applies to
any one of the same-charged leptons,
i.e., $\ell^+ \ell^+ = e^+e^+, \mu^+ \mu^+$ or $\tau^+ \tau^+$,
when the top-quarks decay leptonically with
$BR(t \to W^+ b \to \ell^+ \nu_\ell b) \sim 0.1$.

Considering the  same-sign dilepton asymmetry defined in Eq.~\ref{asym},
in the up-type LQ case we find that
${\cal A}_{\tau \tau}$ may be detected with a statistical significance of
$N_{SD} \sim 8$, with an integrated luminosity of
300 inverse fb and a combined efficiency of $\epsilon \sim 0.01$ (see Eq.~\ref{NSD}). On the other hand,
a statistically significant signal of the asymmetries
${\cal A}_{ee/\mu\mu}$ will require the 13 TeV HL-LHC with
an integrated luminosity of
3000 inverse fb.

\subsection{Expectations at higher energy hadron colliders}
\begin{figure}[htb]
\begin{center}
\includegraphics[scale=0.55]{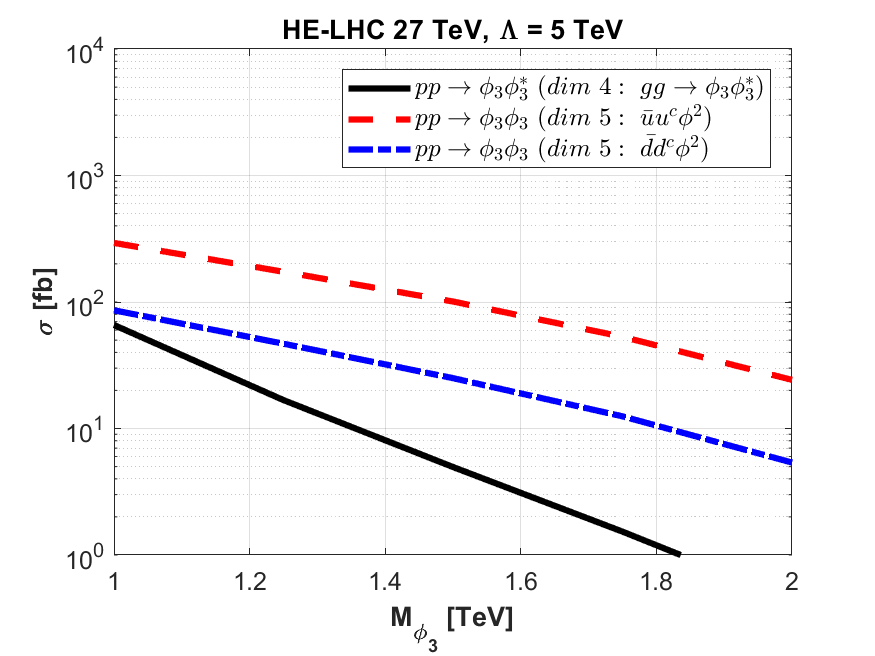}
\includegraphics[scale=0.55]{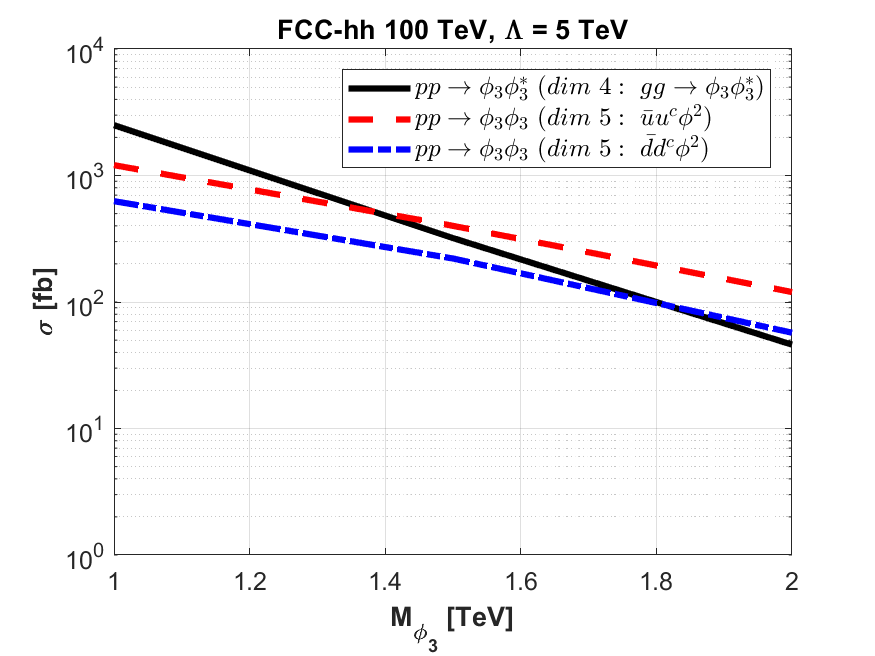}
\end{center}
\vspace{-0.8cm}
\caption{ Pair-production cross-sections of the down-type
and up-type LQ, as a function of the LQ mass, for a NP scale
$\Lambda=5$ TeV, at a 27 TeV HE-LHC (upper plot) and
a 100 TeV FCC-hh (lower plot): the QCD cross-section
via $gg, q \bar q \to \phi_3 \phi_3^*$ (solid line),
the same-charge up-type LQ pair-production cross-section
via $u u \to \phi_3 \phi_3$ (dashed line) and
the same-charge down-type LQ pair-production cross-section
via $dd \to \phi_3 \phi_3$ (dashed-dotted line).
See also text.}
\label{LQCSX5TeV}
\end{figure}
\begin{figure}[htb]
\begin{center}
\includegraphics[scale=0.55]{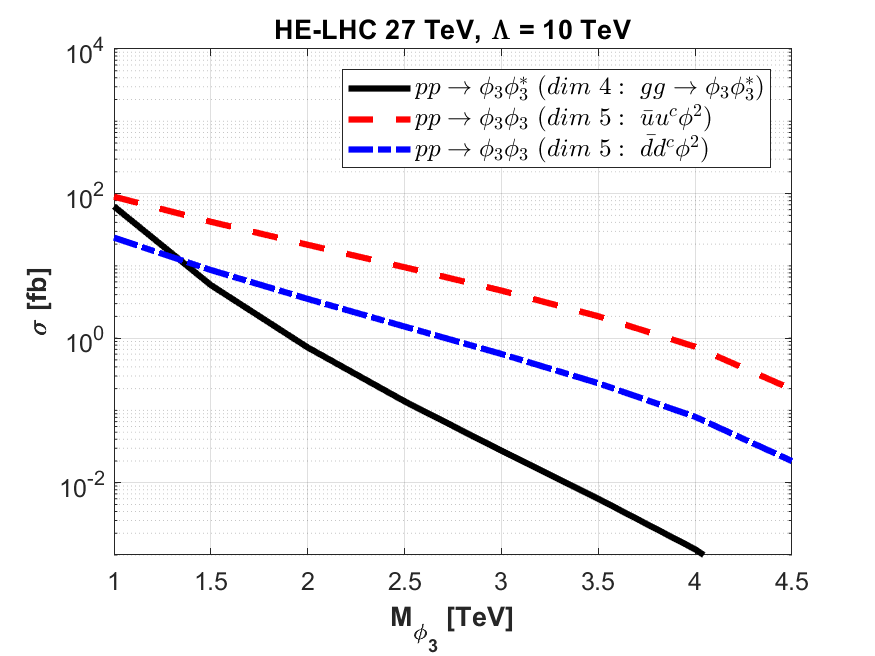}
\includegraphics[scale=0.55]{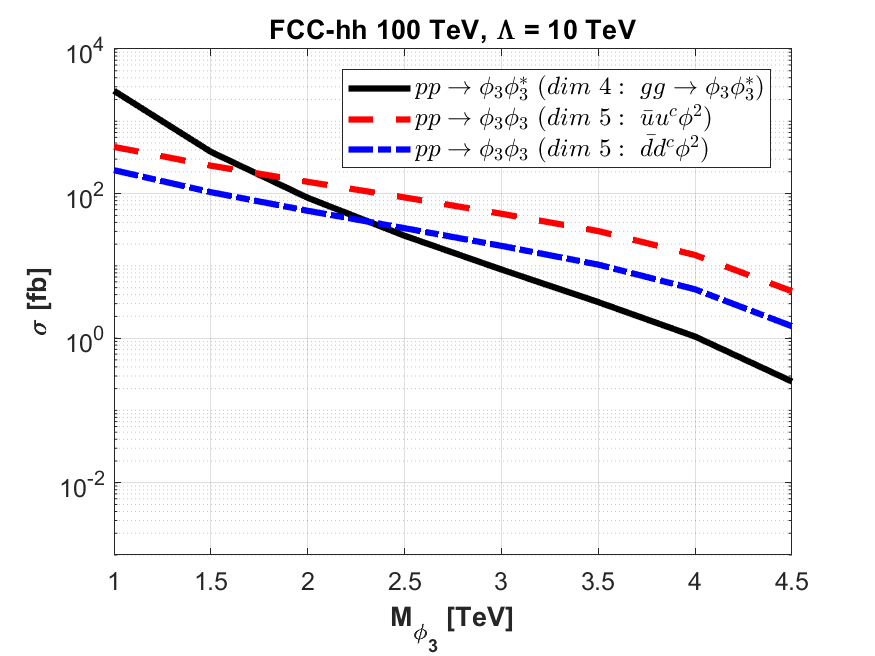}
\end{center}
\vspace{-0.8cm}
\caption{Same as Fig.~\ref{LQCSX5TeV} for $\Lambda=10$ TeV.}
\label{LQCSX10TeV}
\end{figure}
\begin{figure}[htb]
\begin{center}
\includegraphics[scale=0.55]{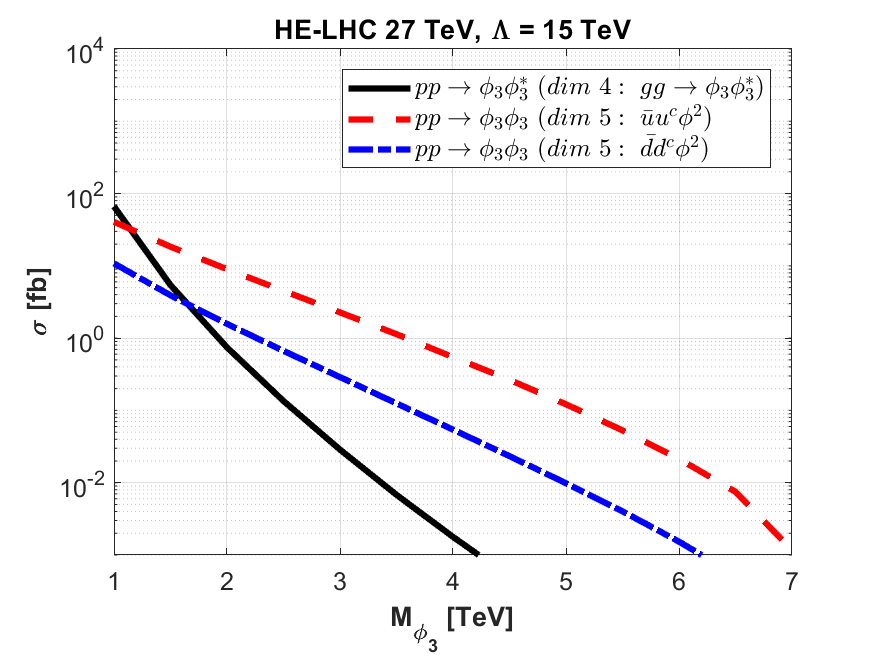}
\includegraphics[scale=0.55]{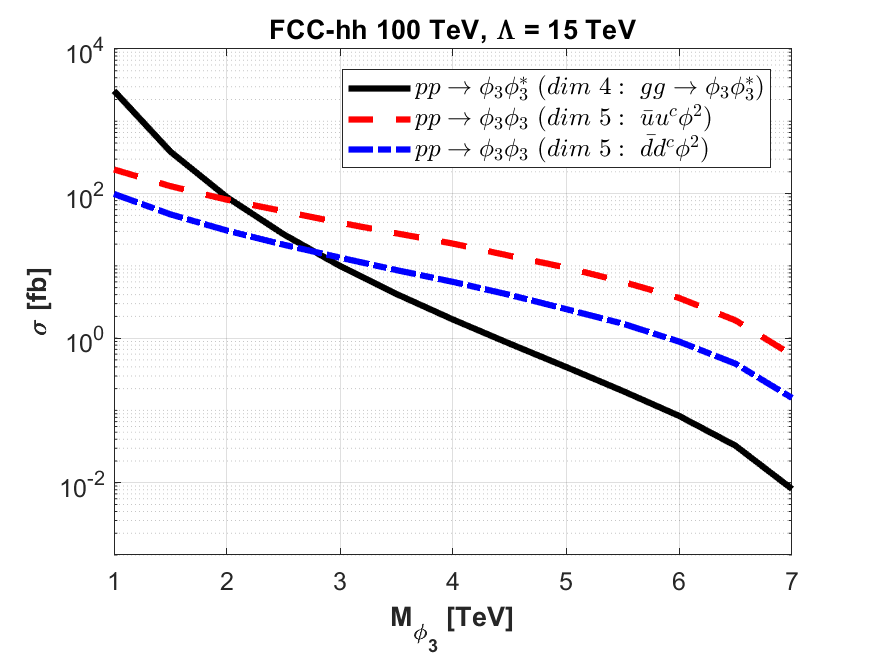}
\end{center}
\vspace{-0.8cm}
\caption{Same as Fig.~\ref{LQCSX5TeV} for $\Lambda=15$ TeV.}
\label{LQCSX15TeV}
\end{figure}

As can be seen from Fig.~\ref{LQCSX},
the LQ production cross-sections
sharply drop with the LQ mass at the 13 TeV LHC
for LQ masses $M_\phi > 1$ TeV.
This is due to the limited phase
space at the 13 TeV LHC for producing TeV-scale
heavy particles and, hence, the currently relatively poor
discovery potential for such new heavy particles.
In particular, the detection of NP
scales $\Lambda > 5$ TeV and/or
heavy new particles with masses of several TeV, will require
in general higher energy colliders with higher luminosities.
For example, for a LQ mass of
$M_{\phi_3} \sim 4$ TeV, the opposite-charge $\phi_3 \phi_3^*$
pair-production cross-section (via $gg,q \bar q \to \phi_3 \phi_3^*$)
at the 13 TeV LHC
is $\sigma(pp \to \phi_3 \phi_3^*) \sim 10^{-6}$ fb.
The new same-charge $\phi_3 \phi_3$ signal discussed above
is also too small
at the 13 TeV LHC for $M_{\phi_3} \sim 4$ TeV;
$\sigma(pp \to \phi_3 \phi_3) \sim 10^{-4}$ fb,
if the NP scale is $\Lambda \sim 10$ TeV.
Therefore, heavy LQ with masses of several TeV are not accessible at the
13 TeV LHC with or without the new EFT interactions from the higher dimensional
effective operators.

A better sensitivity to multi-TeV LQ and, in particular, to the LQ EFT dynamics presented in this work, can be
obtained at future higher energy hadron colliders such as
the HE-LHC and the FCC-hh mentioned above.
In Figs.~\ref{LQCSX5TeV}-\ref{LQCSX15TeV} we plot
the same-charge LQ pair-production cross-sections
$pp \to \phi_3 \phi_3 $ for both the down-type and up-type LQ
(i.e., the underlying hard-processes being $dd \to \phi_3 \phi_3$
and $uu \to \phi_3 \phi_3$, respectively), as well
as the opposite-charge LQ pair-production
(QCD) cross-section $pp \to \phi_3 \phi_3^*$
(via $gg, q \bar q \to \phi_3 \phi_3^*$),
for a NP scale of $\Lambda=5,10$ and $15$ TeV.
Here also, for consistency with the EFT framework,
all cross-sections are calculated with an
invariant mass cut on the LQ pair $M_{\phi_3 \phi_3} < \Lambda$, i.e.,
$M_{\phi_3 \phi_3} < 5,10,15$ TeV for $\Lambda=5,10,15$ TeV, respectively.
We note that the cross sections in Figs.~\ref{LQCSX5TeV}-\ref{LQCSX15TeV}
for a 3rd generation LQ are insensitive to the Yukawa couplings in Eq. \ref{phiSMY},
so the results for 1st and 2nd generation LQ are expected to be comparable.

We see that the production rate of positively-charged
up-type LQ pair (in the EFT framework)
can reach $\sigma(pp \to \phi_3 \phi_3) \sim {\cal O}(1)$ fb
at the 100 TeV FCC-hh, for a rather heavy LQ with $M_{\phi_3} \sim 7$ TeV and
a NP scale of $\Lambda \sim 15$ TeV,
whereas the corresponding opposite-charged $\phi_3 \phi_3^*$
signal (i.e., for $M_{\phi_3} \sim 7$ TeV)
is expected to be about two orders of magnitudes smaller.
A 27 TeV HE-LHC is also sensitive to a several TeV LQ and a NP scale of
${\cal O}(10)$ TeV, e.g., expecting an ${\cal O}(1)$ fb
cross-section for pair production of positively-charged
up-type LQ pair when $M_{\phi_3} \sim 4$ TeV and
a NP scale of $\Lambda \sim 10$ TeV.

\section{Summary \label{sec6}}

We have explored the phenomenology
of the EFT expansion of a low-energy TeV-scale framework, where the
``light" degrees of freedom contain the SM fields and a
down-type scalar LQ $\phi(3,1,-\frac{1}{3})$ or an up-type LQ $\phi(3,1,\frac{2}{3})$.

We found that there are only two dimension five operators that can be
assigned to the down-type LQ $\phi(3,1,-\frac{1}{3})$ and
four dimension five operators for the up-type LQ $\phi(3,1,\frac{2}{3})$; all these
dimension five operators
violate lepton number by two units. We have also
identified the distinct underlying heavy
physics that can generate these operators at tree-level.

We have shown that these dimension five operators
can generate sub-eV Majorana neutrino masses at 1-loop and 2-loops,
where, in the 2-loops case, the effective NP scale can be
as low as $[\Lambda/{\rm TeV}]/f \sim 5$,
where $f$ is
the corresponding Wilson coefficient derived from the underlying heavy theory.
We also found that the dimension five operator involving the down-type LQ, $\bar d d^c \phi^2$,   
which is relevant to current collider phenomenology, may mediate neutrinoless double beta decay.

We have then focused on
collider phenomenology of both the down and up-type scalar LQ in the EFT framework.
In particular,
motivated by the current anomalies in B-decays, we
have suggested an approximate $Z_3$ generation symmetry and
studied the signals of 3rd generation down-type and
up-type LQs ($\phi_3$) at the LHC.
We found that
the dimension five operators may give rise to striking asymmetric,
same-charge dilepton final states in the reactions $pp \to \phi_3 \phi_3$
for both the down and up-type scalar LQs, that have
low background.

For example, for the 3rd generation down-type LQ with a mass
$M_{\phi_3} \sim 1$ TeV and a NP scale
$\Lambda \sim 5$ TeV,
the resulting same-sign lepton signature is
$pp \to \phi_3 \phi_3 \to \tau^- \tau^- + 4 \cdot j + 2 \cdot j_b$ 
($j$=light jet and $j_b$=b-jet), which is expected
to yield about 500 such $\tau^- \tau^-$ events at the 13 TeV LHC with a 
luminosity of 300  fb$^{-1}$.
For the 3rd generation up-type LQ,
we expect about 6000 events of same-sign positively charged $\tau^+ \tau^+$ from the process
$pp \to \phi_3 \phi_3  \to \tau^+ \tau^+  + 2 \cdot j_b$,
if $\Lambda \sim 5$ TeV.
Moreover, for similar parameters, the same-charge up-type $\phi_3 \phi_3$ pair production process
can also generate events with pairs of same-charge top quarks
$pp \to tt + \missET$ (when each LQ decays via $\phi_3 \to t \nu$),
leading to about 50 same-sign dilepton events
$pp \to \ell^+ \ell^+  + 2 \cdot j_b + \missET$  (when each top-quark decays leptonically via $t \to W^+ b \to \ell^+ \nu_\ell b$),
for any of the three charged leptons,
$\ell = e,\mu, \tau$.

We have also defined a double lepton-charge asymmetry that may be useful for detection and disentangling these same-sign lepton signals.

Finally, since the LQ production cross-sections
sharply drop with the LQ mass at the 13 TeV LHC,
due to its limited phase-space
for producing multi-TeV
heavy particles, we have also calculated the
projected same-charge LQ pair production cross-sections,
$\sigma(pp \to \phi_3 \phi_3)$, at 27 and 100 TeV hadron colliders;
the future planned HE-LHC and FCC-hh, respectively.
As expected, we find that these future higher energy hadron colliders
can extend the sensitivity to the LQ EFT dynamics
up to masses of $M_{\phi} \gsim 5$ TeV and a NP scale
of $\Lambda \sim 15$ TeV.

\bigskip
\bigskip

{\bf Acknowledgments:}
The work of AS was supported in part by the US DOE contract \#DE-SC0012704.
We also thank the referee for his comment on neutrino masses.

\bigskip
\bigskip

\appendix{{\bf Appendix: Dimension six operators for the down-type scalar LQ $\boldsymbol{\phi(3,1,-\frac{1}{3}}$)} }

\medskip

There are several
classes of dimension six operators which correspond to the generic form of
Eq.~\ref{gen1}, which will be listed here.

The only $\phi^6$ operator is:
\begin{eqnarray}
O_{\phi^6}^{(6)} = \left(\phi^* \phi \right)^3 ~.
\end{eqnarray}

There are no operators of the form $\phi^5 H$ due to gauge invariance and out of the $\phi^4 H^b D^{2-b}$ type operators
there are only two non-redundant gauge invariant operators corresponding to the $b=0$ and $b=2$ cases:
\begin{eqnarray}
O_{\phi^4 H^2}^{(6)} =
\left( H^\dagger H \right) \left( \phi^* \phi \right)^2
~,~ O_{\phi^4 D^2}^{(6)} = |\phi|^2 |D\phi|^2 ~.
\end{eqnarray}

Out of the operators that contain a $\phi^3$ factor,
the ones of the form $\phi^3 H^b D^{3-b}$ are absent since they
violate either gauge ($b$ odd) or Lorentz ($b$ even) invariance.
On the other hand, in the class $\phi^3 \psi^2$ operators
there are four gauge invariant combinations which can be constructed,
all of the form $|\phi|^2 \phi \bar\psi_L \psi_R$:
\begin{widetext}
\begin{eqnarray}
O_{\phi^3 \psi^2}^{(6)} \in
|\phi|^2 \left(\overline q l^c \phi \right) ~,~
|\phi|^2 \left(\overline u e^c \phi \right) ~,~
|\phi|^2 \left(\overline{q^c} q \phi \right) ~,~
|\phi|^2 \left(\overline{u^c} d \phi \right ) ~,
\end{eqnarray}
\end{widetext}
where the last two $\phi^3 \psi^2$ operators above violate both baryon and lepton number.

The operators that contain a $\phi^2$ factor can be divided into two categories: the ones proportional to gauge invariant factor $|\phi|^2$ and the ones that contain $\phi^2$ or $(\phi^*)^2$.
The former case is straight forward, since it includes all operators involving
an SU(3) singlet $\phi^\dagger \phi$ of the form:
\begin{eqnarray}
O_{\phi^2 SM^4}^{(6)} \in
|\phi|^2 {\cal O}_{SM}^4 ~,
\end{eqnarray}
where ${\cal O}_{SM}^4$ includes all the dimension 4 renormalizable terms of the SM
Lagrangian. In addition, there are operators involving the SU(3) octet $\phi^\dagger \phi$
states of the form:
\begin{eqnarray}
\left( \phi^\dagger \lambda^a D_\mu \phi \right)
\left( \bar q \lambda^a \gamma^\mu q \right)~,~
\left( \phi^\dagger \lambda^a \phi \right)
B^{\mu \nu} G^a_{\mu \nu} ~,
\end{eqnarray}
where $\lambda^a$ are the SU(3) Gell-Mann matrices and $B^{\mu \nu}$ is the SM SU(1) field strength.

The latter case (i.e., operators which contain a $\phi^2$ factor) is more elaborate,
but it can be shown that there are only two  non-redundant gauge invariant operators
of this class,
both in the form
$\phi^2 \psi^2 D$, where $\psi^2$ is composed out of
one quark and one lepton:
\begin{eqnarray}
O_{\phi^2 \psi^2 D}^{(6)} \in
\epsilon^{abc} \phi_a \left(D_\mu \phi \right)_b \bar\ell \gamma^\mu q_c ~,~
\epsilon^{abc} \phi_a \left(D_\mu \phi \right)_b \bar e \gamma^\mu d_c ~, \nonumber \\
\end{eqnarray}
where here $a,b,c$ are color indices.

Finally, the dimension six operator which contain only one LQ
field have to be of the form $\phi \psi^2 H^b D^{2-b}$,
where $0 \leq b \leq 2$ and $\psi^2$ is either a quark-lepton or quark-quark pair. For the $b=2$($b=1$) case we find six(five) gauge invariant operators:
\begin{widetext}
\begin{eqnarray}
O_{\phi \psi^2 H^2}^{(6)} &\in&
|H|^2 \phi^\dagger \overline q q^c ~,~
|H|^2 \phi \overline q l^c ~,~
|H|^2 \phi \overline{d^c} u ~,~
|H|^2 \phi \overline u e^c ~,~
 \phi^\dagger (\overline q H)(H^\dagger q^c) ~,~
 \phi (\overline q H)(H^\dagger l^c) ~, \\
O_{\phi \psi^2 H D}^{(6)} &\in&
(\overline q H)\gamma^\mu u^c D_\mu \phi^\dagger ~,~
(\overline q \tilde H)\gamma^\mu d^c D_\mu \phi^\dagger ~,~
(\overline q \tilde H)\gamma^\mu e^c D_\mu \phi ~,~
(\overline l H)\gamma^\mu u^c D_\mu \phi ~,~
(\overline l \tilde H)\gamma^\mu d^c D_\mu \phi ~,
\end{eqnarray}
\end{widetext}
where we have omitted the color indices and the antisymmetric tensor
$\epsilon_{abc}$ in the above operators containing
$3 \otimes 3 \otimes 3 $ and $\overline3 \otimes \overline3 \otimes \overline 3$
states.

The case of $b=0$, i.e., operators of the type $\phi \psi^2 D^{2}$, contain four possible combinations of $\psi^2$ fields
of the form:
\begin{widetext}
\begin{eqnarray}
O_{\phi \psi^2 D^2}^{(6)} \in
D^2 \times \overline q q^c \phi^* ~,~
D^2 \times \overline q l^c \phi   ~,~
D^2 \times \overline{d^c} u \phi   ~,~
D^2 \times \overline u e^c \phi   ~,
\end{eqnarray}
\end{widetext}
where the notation above indicates that the two derivatives are to act on any of the fields;
note though that $D_\mu D^\mu$ acting on a field gives a redundant operator, but
$[D_\mu ,\,D_\nu]$ does not. Thus, for example,
$ D^2\times \overline q l^c \phi^\dagger $ corresponds to:
\begin{widetext}
\begin{eqnarray}
D^2 \times \overline q l^c \phi  \to
(\overline q D_\mu l^c) D^\mu \phi ~,~
(\overline q \sigma_{\mu\nu} l^c) B^{\mu\nu} \phi ~,~
(\overline q \sigma_{\mu\nu} \sigma^I l^c) W_I^{\mu\nu}\phi ~,~
(\overline q \sigma_{\mu\nu} \lambda^A l^c) G_A^{\mu\nu}\phi ~.
\end{eqnarray}
\end{widetext}

\end{document}